\PassOptionsToPackage{numbers,square,sort&compress,comma}{natbib}
\documentclass[preprint,5p,times,10pt,twocolumn]{elsarticle}

\usepackage{amssymb}
\usepackage{amsmath}

\usepackage{tablefootnote}
\usepackage{subcaption}
\usepackage{siunitx}
\usepackage[super]{nth}
\usepackage{threeparttable}
\usepackage{subcaption}
\usepackage{gensymb}
\usepackage{color}
\usepackage{cuted}
\usepackage{multicol}
\usepackage[normalem]{ulem} 
\usepackage{hyperref}
\usepackage{booktabs}
\usepackage{scrextend}
\usepackage{xcolor,soul}
\usepackage[table]{xcolor}
\usepackage{multirow}
\usepackage{tabularx}
\usepackage{upgreek}

\renewcommand{\vec}[1]{\boldsymbol{\mathrm{#1}}}

\definecolor{RED}{rgb}{1,0,0} 
\definecolor{BLUE}{rgb}{0,0,1} 
\definecolor{GREEN}{rgb}{0,0,1}

\newcommand{\U}[1]{[\makebox[2.5em][c]{#1}]}

\newcommand{\rmmu}{\upmu}
\newcommand{\tsup}[1]{\textsuperscript{#1}}
\newcommand{\tmin}{\textminus}

\journal{Acta Astronautica}

\begin{document}

\begin{frontmatter}



\title{Feasibility of Capillary-Driven Orbital Liquid Mirror Telescopes}

%

\author[1]{Janoah {Dietrich}\corref{cor1}%
\fnref{fn1}}
\ead{janoah.dietrich@gatech.edu}

\author[1]{{\'A}lvaro {Romero-Calvo}
\fnref{fn2}}
\ead{alvaro.romerocalvo@gatech.edu}

\cortext[cor1]{Corresponding author.}

\affiliation[1]{organization={Daniel Guggenheim School of Aerospace Engineering, Georgia Institute of Technology},
                addressline={620 Cherry Street NW}, 
                city={Atlanta},
                postcode={30332}, 
                tate={GA},
                country={United States}}

\begin{abstract}
Space-based liquid mirror telescopes (LMTs) have been recently proposed to overcome the scaling limits of traditional rigid mirrors. In these new space architectures, capillary forces would shape a reflective liquid into a spherical optical surface in microgravity. However, the equilibrium, stability, and dynamic behavior of such liquid interfaces remain largely uncharacterized under orbital perturbations. This study investigates the feasibility of space LMTs by combining axisymmetric capillary with thin-film models. Solar radiation pressure and self-gravitation are found to be the dominant perturbations in Sun--Earth L2 orbits, causing equilibrium deformations that exceed the $\lambda/20$ optical threshold at apertures above 0.86~m for visible ($\lambda = 380$~nm) and 1.58~m for near-infrared observatories ($\lambda = 2.5~\rmmu \mathrm{m}$). Micrometeoroid impacts self-heal within seconds to minutes, settling after filling requires years to decades at large apertures, and propulsive maneuvers breach optical limits within seconds to hours. These results highlight the need for interface control in large-scale systems such as NASA's 50~m Fluidic Telescope (FLUTE). Thermocapillary actuation supports apertures of 2.3/8.0~m (VIS/NIR) at $100~\rmmu \mathrm{K}$ and 10.6/37.7~m at $10~\rmmu \mathrm{K}$ resolution. Simultaneously, parasitic Marangoni flows impose sub-mK uniformity requirements for NIR and 1--$100~\rmmu \mathrm{K}$ for visible operation on any capillary LMT. Departing the capillary regime by imposing a surface-normal settling acceleration, ranging from 39~mm\,s\tsup{\tmin2} (NIR,5~m) to 17~m\,s\tsup{\tmin2} (VIS,50~m), is the most promising route to tens of meters in aperture. This approach enables a passive implementation, effectively suppressing perturbation induced interface distortions and thermal Marangoni flows.
\end{abstract}


\begin{highlights}
\item Capillary liquid mirrors may enable low-cost, large-aperture space observatories.
\item Orbital perturbations deform capillary mirrors beyond optical limits at 0.9--1.6~m aperture.
\item Interface control is required both during assembly and steady-state operation.
\item Thermocapillary actuation extends viable apertures to several meters.
\item Surface-normal settling accelerations present a path to tens-of-meter apertures.
\end{highlights}


\end{frontmatter}



\section{Introduction}
Pushing the boundaries of observable space and resolving ever-fainter cosmic phenomena remain key drivers in the development of astronomical instrumentation. The advancement of telescope technologies has enabled the measurement of cosmological expansion with unprecedented accuracy~\cite{Riess2022, Planck2020}, facilitated the structural analysis of distant galaxies~\cite{Koekemoer2011, Shapley2011, Scoville2007}, and allowed for the investigation of high-energy cosmic phenomena that govern the evolution of the universe~\cite{Meszaros2006, Wilms2001, Gardner2006}, among many others~\cite{photonics12030199, Marov2015}.

\begin{figure*}[t]
    \centering
    \includegraphics[width=\textwidth]{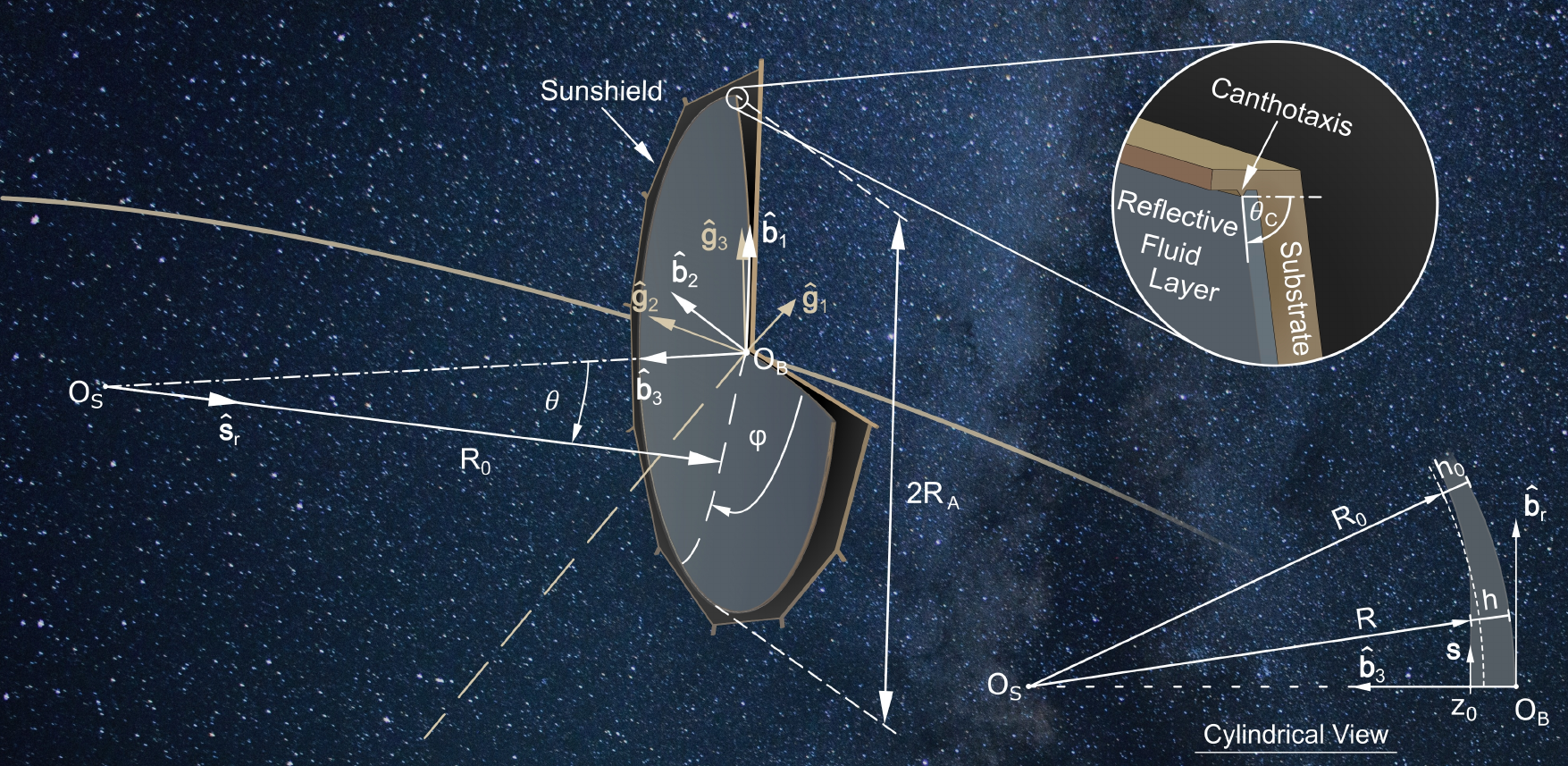}
    \caption{Conceptual illustration of an orbital liquid mirror telescope comprising a rigid spherical substrate filled with a reflective vacuum-compatible liquid and a sunshield. Four coordinate systems are introduced: the inertial orbital frame $\mathcal{G} = \{ O_B,\vec{\hat{g}}_1,\vec{\hat{g}}_2,\vec{\hat{g}}_3 \}$, the telescope-fixed Cartesian frame $\mathcal{B}_\text{Cart} = \{ O_B,\vec{\hat{b}}_1,\vec{\hat{b}}_2,\vec{\hat{b}}_3 \}$, the telescope-fixed cylindrical frame $\mathcal{B} = \{ O_B,\vec{\hat{b}}_r,\vec{\hat{b}}_\phi,\vec{\hat{b}}_3 \}$, and the surface-following spherical frame $\mathcal{S} = \{ O_S,\vec{\hat{s}}_r,\vec{\hat{s}}_\theta,\vec{\hat{s}}_\phi \}$. Their respective roles are defined in Sec.~\ref{sec:methods/problem_Statement}.}
    \label{fig:flute}
\end{figure*}

This fundamental scientific drive motivates the development of telescopes with larger primary mirror apertures~\cite{Nasa_Strategic_Plan}, since the telescope resolution and contrast are directly limited by its light-gathering capability~\cite{postman2012advanced}. Although Earth-based facilities continue to expand, atmospheric scattering, absorption, refraction, and turbulence impose fundamental performance limitations~\cite{English2017,postman2012advanced}. These effects degrade image quality, reduce the signal-to-noise ratio, and constrain the observable spectral wavelengths~\cite{BahcallSpitzer1982,English2017}. As a result, the trend toward larger apertures favors space-based observatories. While space-based telescopes are free from atmospheric interference~\cite{postman2012advanced}, they are constrained by launch-vehicle fairing limitations on volume, mass, and rigidity. Deployable structures, such as those used on the James Webb Space Telescope (JWST)~\cite{nasaDeploymentExplorer}, can help overcome these issues. However, their cost scales approximately quadratically with size, imposing significant financial burdens on new missions~\cite{Stahl2010Survey}. 

These constraints motivate the exploration of alternative architectures for large-aperture mirrors. Liquid mirror telescopes (LMTs), which use fluid interfaces to form optically smooth reflective surfaces, have been proposed as a path forward. The LMT concept originates from Newton’s rotating bucket experiment~\cite{Newton1687}, where a spinning liquid in a gravitational field naturally forms a paraboloid, making it an ideal candidate for a primary mirror due to its inherently smooth surface. Earth-based LMTs date back to the 1850s~\cite{Duncan1986}, with Wood~\cite{Wood1909} providing the first detailed prototype description. The concept gained broader recognition with Borra in the late \nth{20} century~\cite{Borra1982,Borra1992,BORRA1985} and was subsequently implemented in the 3~m NASA Orbital Debris Observatory~\cite{Potter1996,Cabanac1998,Hebert2004}, the 6~m Large Zenith Telescope~\cite{Hickson2007,Hickson2007a}, and the 4~m International Liquid Mirror Telescope~\cite{Surdej2018,Pradhan2019}. Proposals for lunar-based LMTs have garnered significant attention~\cite{Borra1991,Angel2008,Schauer2020,Detsis2013,Comstock2025}, and a collaboration between NASA and the Technion–Israel Institute of Technology is currently addressing the feasibility of the space-based 50-meter FLUidic TElescope (FLUTE)~\cite{Gregg2024_FluidicMirrorSupportStructure}. In this proposal, the shape of the mirror is driven by surface tension rather than the combination of gravitational and centripetal forces. Thus, the interface is believed to be spherical in microgravity, making corrective optics necessary for operation~\cite{Gabay2025_FluidDynamicsLiquidMirror,Flute2}.

The fluidic nature of FLUTE-like telescopes may lead to several benefits, including independence from expensive machining and polishing~\cite{Wu2013_MFDM_Book,zenith_overview_2024}, self-healing capabilities~\cite{Zheng2018_HealingCapillaryFilms}, and simplified in-space assembly driven by liquid wetting~\cite{Gregg2024_FluidicMirrorSupportStructure,Rowlands2024_SelfAssemblingILM}. However, these same properties make LMTs inherently sensitive to external perturbations, which can distort the mirror surface from the spherical shape and compromise optical performance~\cite{Myshkis1987, Collicott2007}. Understanding the equilibrium, stability, and dynamic response of such fluidic interfaces is therefore key to assessing the feasibility and limitations of the concept.

In orbit, spacecraft are continuously exposed to a variety of environmental forces and disturbances, including solar radiation pressure (SRP) \cite{McInnes1999}, tidal effects \cite{Schaub2018}, self-gravitation \cite{Collicott2007, Myshkis1987}, and g-jitter \cite{Antar}. For conventional rigid spacecraft structures, these effects are typically negligible because their magnitudes are small per unit mass or area. In contrast, large, lightweight, or in-space-manufactured structures are far more susceptible to such influences~\cite{Pappa2003}. This sensitivity becomes critical for large-scale fluidic systems, such as LMTs, whose equilibrium interfaces exist in a regime dominated by extremely small restoring forces. As the characteristic length of a liquid mirror increases, capillary forces diminish relative to external accelerations due to the scaling of the Bond number with $\propto L^{2}$, amplifying the impact of even small perturbations. Despite the theoretical promise of fluidic telescopes to produce spherical profiles, the optical interface response to spacecraft disturbances remains poorly investigated. Consequently, establishing the feasibility of large-aperture fluid observatories requires quantifying this sensitivity. 

This paper evaluates the feasibility and scaling limits of capillary-driven LMTs, focusing on their susceptibility to perturbations in representative orbits and determining the conditions under which high-fidelity optical performance can be preserved. Section~\ref{sec:methods} introduces the system parameters of the orbital LMT study, characterizes the dominant orbital perturbations, and presents the governing equations for liquid interface equilibrium and dynamics together with their numerical solution procedure. In Section~\ref{sec:resdis}, the resulting equilibrium interfaces are analyzed, and aperture limits bounding the optically compatible regime are determined. The dynamic response to perturbations is assessed using a numerical thin-film model that covers settling behavior during late-stage filling, recovery following particle impacts, and responses to external accelerations. These results inform a discussion on mitigation strategies in which different interface control mechanisms are evaluated.

\section{Methods}\label{sec:methods}

\subsection{Problem Statement}\label{sec:methods/problem_Statement}
The subsequent analysis focuses on orbital liquid-mirror telescopes characterized by a thin liquid film supported by a spherical dish with $2R_A \in [1,50]$~m aperture. With the paraxial focal length of a spherical mirror given by $f = R_0/2$ and the focal ratio defined as $F = f/(2R_A)$~\cite{Cheng2009}, the reflective surface has a radius of curvature $R_0 = 4FR_A$. A radius of curvature of the dish $R_D = R_0 + h_0$ is therefore adopted, where $h_0 \in [1,20]$~mm denotes the nominal liquid film thickness and $F = 1$ the focal ratio. Four coordinate systems are used within the analysis. Frames $\mathcal{B} = \{ O_B,\vec{\hat{b}}_r,\phi,\vec{\hat{b}}_3 \}$ and $\mathcal{S} = \{ O_S,\vec{\hat{s}}_r,\theta,\phi \}$ are the telescope-fixed frames used to describe perturbations and the liquid interface in cylindrical and spherical coordinates, respectively. The telescope-fixed cartesian frame $\mathcal{B}_\text{Cart} = \{ O_B,\vec{\hat{b}}_1,\vec{\hat{b}}_2,\vec{\hat{b}}_3 \}$ and the orbital Hill frame $\mathcal{G} = \{ O_B,\vec{\hat{g}}_1,\vec{\hat{g}}_2,\vec{\hat{g}}_3 \}$ are employed in the derivation of the tidal accelerations~\cite{Schaub2018}. The local film thickness is defined in $\mathcal{S}$ as ${}^{\mathcal{S}}h(\phi, \theta) = R_D - r_S(\phi, \theta)$. Thickness perturbations $\delta h = h - h_0$ represent deviations of the optical surface from an ideal spherical cap of radius $R_0 = 4R_A$. In frame $\mathcal{B}$, where $r$ and $z$ denote the coordinates along $\vec{\hat{b}}_r$ and $\vec{\hat{b}}_3$, respectively, this surface follows the form
\begin{equation}
    z(r) = R_0 + z_0 - \sqrt{R_0^2 - r^2},
\end{equation}
terminating at height $z_e$. As illustrated in Fig.~\ref{fig:flute}, the liquid interface is pinned at the rim via canthotaxis, allowing a range of interface slopes at a fixed contact line~\cite{Langbein2002}. 

Galinstan is adopted as the baseline mirror liquid, with generalizations to other materials discussed where applicable. This choice follows previous research identifying Galinstan as a particularly favorable candidate for liquid mirror applications~\cite{Dietrich_RomeroCalvo_2026}. At an operating temperature of $T_0 = -10$\,°C, Galinstan is characterized by density $\rho_L = 6440$~kg\,m$^{-3}$, dynamic viscosity $\mu = 2.4 \times 10^{-3}$~Pa\,s, surface tension $\gamma_0 = 718$~mN\,m$^{-1}$, and surface tension temperature coefficient $\gamma_T = -1.09 \times 10^{-5}$~N\,m$^{-1}$\,K$^{-1}$~\cite{HandschuhWang2021,Kramer2014}\footnote{Literature values for the surface tension of Galinstan range from 535 to 718~mN\,m$^{-1}$, with variability attributed to differences in alloy composition and oxidation state~\cite{HandschuhWang2021,Kramer2014}. The upper bound is adopted to assess mirror feasibility under the most favorable capillary conditions.}. The analysis assumes a Sun–Earth L2 (SEL2) orbit to avoid the severe drag and tidal perturbations that impact large spacecraft in Earth orbits.

In the absence of a well-established scaling law for the mass distribution of orbital LMTs, a first-order scaling law is adopted with the form
\begin{equation}\label{eq:mass}
    m_{\text{SC}}(R_A)
    =
    \exp\left({\frac{ \dot{\Delta v}\, T_{\text{m}} }{I_{sp} \, g_0}} \right)
    \, m_{\text{dry}}(R_A),
\end{equation}
\begin{equation}\textbf{}
    m_{\text{dry}}(R_A) = \pi R_A^2 h \left(\rho_L + \frac{1}{2} \rho_{\text{S}}\right) + m_{\text{bus,dry}}, 
\end{equation}
where $\dot{\Delta v} = \dot{\Delta v}_{SK,\text{JWST}}  \,  (\pi R_A^2 / A_{\text{JWST}}) \, (m_{\text{JWST}}/m_{\text{dry}})$ is the annual stationkeeping budget, dimensioned by the budget for JWST $\dot{\Delta v}_{SK,\text{JWST}} \approx 2.43$~m\,s$^{-1}$\,yr$^{-1}$~\cite{Dichmann2014} with the JWST sunshield area $A_{\text{JSWT}} = 140$~m$^2$, the JWST operational mass $m_{\text{JWST}} \approx 6200$~kg$^{\ref{fn:nasa}}$, and the mission lifetime $T_{\text{m}} \approx 10$~yr. The ratio $(\pi R_A^2 / A_{\text{SS}})(m_{\text{JWST}}/m_{\text{dry}})$ scales the JWST stationkeeping budget by the SRP acceleration ratio, assuming equivalent SEL2 halo orbit geometry and reflectivity. Due to the small propellant fraction ($\Delta v_{\text{tot}} \ll I_{sp} g_0$), $m_{\text{SC}} \approx m_{\text{dry}}$ is used to evaluate the propellant mass. $m_{\text{bus,dry}} \approx 5200$~kg is the mass of all components that do not scale with the mirror radius, assumed to be equivalent to JWST without its mirror and propellant\footnote{\label{fn:nasa}\url{https://science.nasa.gov/mission/webb/fact-sheet/}. Consulted on: 23/03/2025}. The mirror assembly consists of a liquid film of density $\rho_L$ and an aluminum dish of density $\rho_S = 2700$~kg\,m$^{-3}$ and structural thickness $h/2$. Propellant is sized using the specific impulse $I_{sp} = 295$~s of the JWST bipropellant SCAT thrusters~\cite{Menzel2023}, with $g_0 = 9.81$~m\,s$^{-2}$. All mass contributions except $m_{\text{mirror}}$ are assumed to be co-located at $r_{\text{B,SB}} = 0$, $z_{\text{B,SB}} = -4$~m.

\subsection{Mathematical Framework}

\subsubsection{Governing Equations}\label{sec:methods/GovEqs}
The evolution of the incompressible, Newtonian liquid mirror interface follows the Navier--Stokes equations~\cite{dreyer_free_2007}
\begin{equation}\label{eq:continuity}
    \nabla \cdot \vec{u} = 0,
\end{equation}
\begin{equation}
    \rho_L\left[
    \frac{\partial \vec{u}}{\partial t}
    + (\vec{u}\cdot\nabla)\vec{u}
    \right]
    = -\nabla p + \mu \nabla^2 \vec{u} + \rho_L \vec{f},
    \label{eq:NS}
\end{equation}
where $\vec{u}$ denotes the velocity field, $p$ the static pressure, $t$ the time, and $\vec{f}$ the body force per unit mass, expressed as the gradient of a scalar mass-force potential $\vec{f} = -\nabla \Pi$ for conservative acceleration fields. These accelerations encompass all external perturbations considered in this work.

\subsubsection{Equilibrium Interface}
\label{sec:methods/EquilibriumInterfaces}

In static equilibrium, the velocity field vanishes and the incompressible Navier-Stokes equations simplify to the Euler condition. After inserting the mass-force potential and accounting for the capillary pressure jump at the interface, the capillary equation~\cite{Myshkis1987}
\begin{equation}\label{eq:Equilibrium}
    \rho_L \Pi = 2 \gamma \mathcal{H}
\end{equation}
is obtained, where $\mathcal{H}$ is the mean interface curvature. The pressure in the surrounding medium is effectively zero due to the vacuum environment, resulting in a small negative pressure that makes the liquid remain under traction~\cite{Caupin2012}.

\subsubsection{Interface Dynamics}
\label{sec:methods/InterfaceDynamics}

In the assessment of the dynamic response, it is convenient to work in frame $\mathcal{S}$ and to introduce the reduced pressure
\begin{equation}\label{eq:modifiedpressure}
    p^* = p + \rho \Pi
\end{equation} 
and the surface fluid velocity perpendicular to $\vec{r_S}$ 
\begin{equation}
    {}^{\mathcal{S}}\vec{u}_{\parallel} = u_{\theta}\,\vec{\hat{s}}_{\theta} + u_\phi\,\vec{\hat{s}}_\phi.
\end{equation} 
Under the assumption of a shallow fluid film ($h \ll R_A$), conservative body forces, and negligible inertia ($Re_{\parallel} \varepsilon^2  = \rho u h^2 / (\mu R_A) \ll 1$), Eq.~\ref{eq:NS} reduces to the lubrication equations~\cite{Oron1997ThinFilms,Leal2007AdvancedTransport,Haskett2005}
\begin{equation}\label{eq:lubrication_equations}
    \mu \frac{\partial^2 \,\, {}^{\mathcal{S}}\vec{u}_{\parallel}}{\partial r_S^2} = \nabla_{\parallel} p^*, \qquad \qquad
    \frac{\partial p^*}{\partial r_S } \approx 0,
\end{equation}
where $r_S$ is the radial distance from $O_\mathcal{S}$ and $\nabla_{\parallel}=~\vec{\hat{s}}_{\theta} R^{-1}\partial_{\theta} + \vec{\hat{s}}_\phi [R\sin(\theta)]^{-1}\partial_\phi$ the surface gradient operator on the sphere, with the local interface radius $R$. Integrating Eq.~\ref{eq:lubrication_equations} twice with respect to $r_S$, applying the no-slip condition at the substrate ($u_{\theta}(r_S=R+h) = u_\phi(r_S=R+h) = 0$), and the no tangential stress condition at the free surface ($\partial u_{\theta}/\partial r_S \big\rvert_{r_S=R} = \partial u_\phi/\partial r_S \big\rvert_{r_S=R} = 0$) yields the velocity profiles~\cite{Oron1997ThinFilms}
\begin{equation}\label{eq:vel_profiles_thinfilm}
    \begin{bmatrix} 
        u_{\theta} \\ u_\phi 
    \end{bmatrix} 
    = \frac{(r_S-R)^2-h^2}{2\mu} \nabla_{\parallel} p^*.
\end{equation}
Integrating the continuity equation (Eq.~\ref{eq:continuity}) across the film thickness $r_S\in[R,R+h]$ and applying the kinematic boundary condition at the free surface, $\partial h/
\partial t=-u_{r_S} (r_S=R)$, yields the surface mass conservation equation~\cite{Oron1997ThinFilms}
\begin{equation}
    \frac{\partial h}{\partial t}
    +
    \nabla_{\parallel} \cdot \vec {}^{\mathcal{S}} \vec q
    =
    0,
    \label{eq:h_conservation}
\end{equation}
where the depth-integrated volumetric flux along the spherical surface is defined as ${}^{\mathcal{S}}\vec q = \int_{R}^{R+h} \,{}^{\mathcal{S}}\vec u_{\parallel}(r_S)\, dr_S.$ After accounting for the velocity profiles in Eq.~\ref{eq:vel_profiles_thinfilm}, Eq.~\ref{eq:h_conservation} becomes
\begin{equation}
    \frac{\partial h}{\partial t}
    =
    \nabla_\parallel \cdot
    \left(
        \frac{h^3}{3\mu}\nabla_\parallel p^*
    \right).
    \label{eq:h_evolution_p}
\end{equation}
To close the system, the pressure at the free surface is related to the interface shape through the normal stress balance. Using the Laplace–Young relation together with the mass-force potential formulation, the reduced pressure (Eq.~\ref{eq:modifiedpressure}) at the free surface becomes
\begin{equation}\label{eq:pstar_surface}
    p^* = -2\gamma\mathcal{H} + \rho \Pi.
\end{equation}
Substituting this expression for $p^*$ into Eq.~\ref{eq:h_evolution_p} yields the thin-film evolution equation
\begin{equation}
    \frac{\partial h}{\partial t}
    =
    \nabla_\parallel \cdot
    \left[
        \frac{h^3}{3\mu}
        \nabla_\parallel
        \left(
            -2\gamma \mathcal{H}
            + \rho \Pi
        \right)
    \right].
    \label{eq:thinfilm_pde}
\end{equation}

\subsubsection{Liquid Interface Control}
\label{sec:methods/ScalingAnalysis}

Any control mechanism aimed at correcting interfacial deformations may enter the governing equations through body forces acting on the liquid or through normal and tangential stresses at the free surface. Conservative body forces can be incorporated into Eq.~\ref{eq:thinfilm_pde} as additive contributions to the mass-force potential, $\Pi_\text{ctrl}$. Non-conservative body forces would require additional forcing terms in the momentum equation. A prescribed normal stress $p_n$, taken positive along the outward surface normal, enters instead through the normal stress balance, so that Eq.~\ref{eq:pstar_surface} becomes~\cite{Rosensweig}

\begin{equation}\label{eq:pstar_normal}
p^* = -2\gamma\mathcal{H} + \rho\Pi - p_n .
\end{equation}

Normal stresses therefore appear in Eq.~\ref{eq:thinfilm_pde} in the same position as the mass-force term $\rho\Pi$, and the two are interchangeable within this formulation. Candidate sources include electrostatic tractions on a conducting or dielectric interface~\cite{MelcherTaylor1969}, magnetic normal stresses on a magnetic fluid~\cite{Rosensweig}, and acoustic radiation pressure~\cite{Borgnis1953,Beyer1978}. Although in each case the same physical mechanism may also generate tangential stresses or bulk flow, which must be assessed separately. Tangential stresses instead drive interfacial flow directly, and can originate from thermal and chemical concentration gradients~\cite{Park2023}.

Among the different alternatives, thermocapillary (Marangoni) actuation can be practically implemented through distributed heat load arrays. The effect enters Eq.~\ref{eq:thinfilm_pde} as a tangential surface stress that results in~\cite{Oron1997ThinFilms,Craster2009}
\begin{equation}
    \frac{\partial h}{\partial t}
    =
    \nabla_\parallel \cdot
    \left[
        \frac{h^3}{3\mu}
        \nabla_\parallel\!\left(-2\gamma\mathcal{H} + \rho\Pi\right)
        -
        \frac{h^2}{2\mu}\,\nabla_\parallel \gamma
    \right].
    \label{eq:thinfilm_marangoni}
\end{equation}

The divergence of the total flux has to vanish ($\nabla_\parallel \cdot\, \vec{q}\big|_{h=h_\text{target}} = 0$) for a stationary interface to form at the target height over the substrate $h_\text{target}(\phi, \theta)$. Modeling the fluid as linearly dependent on the temperature~\cite{Plevachuk2014}
\begin{equation}
    \gamma = \gamma_0 + \gamma_T(T - T_0),
\end{equation}
and evaluating at $h = h_\text{target}$, with the constant target curvature $\mathcal{H}_\text{target}$ of the spherical profile, the steady-state condition yields an elliptic partial differential equation for the unknown temperature field $T(\theta,\phi)$,
\begin{equation}
    \nabla_\parallel \cdot \bigl( A\, \nabla_\parallel T \bigr)
    =
    \frac{1}{\gamma_T}\nabla_\parallel \cdot \bigl( D\, \nabla_\parallel f \bigr),
    \label{eq:elliptic_pde}
\end{equation}
where $f = -2(\gamma_0 - \gamma_T T_0)\mathcal{H}_\text{target} + \rho\Pi_\text{dist}$ is the known driving function and

\begin{equation}
    D = \frac{h_\text{target}^3}{3\mu}, \qquad
    A = \frac{2\mathcal{H}_\text{target}\, h_\text{target}^3}{3\mu}
        + \frac{h_\text{target}^2}{2\mu}.
    \label{eq:DA_coeffs}
\end{equation}
Equation~\ref{eq:elliptic_pde} is subject to regularity at $\theta = 0$ and a no-flux condition at the pinned contact line $\theta = \theta_C = \arcsin{\left( R_A/R \right)} \approx 14.48$°. The minimum-energy solution is obtained by enforcing zero flux pointwise, reducing Eq.~\ref{eq:elliptic_pde} to $A\,\nabla_\parallel T = D/\gamma_T\,\nabla_\parallel f$. In the thin-film limit $\mathcal{H}_\text{target} h_\text{target} \ll 1$, the ratio $D/A \approx 2h_\text{target}/3$, and the required temperature gradient simplifies to

\begin{equation}
    \nabla_\parallel T_\text{req}
    =
    \frac{2h_\text{target}}{3\gamma_T}
    \,\nabla_\parallel f.
    \label{eq:T_req}
\end{equation}

\subsection{Perturbations}
\label{sec:methods/Perturbations}

Spacecraft are exposed to a wide range of environmental factors that can and do result in dynamic perturbations. This section includes a selection of the most significant effects in a priori compatible orbits beyond LEO. For comparing the relative effect of these perturbations on the interface, the non-dimensional Bond and centripetal Bond numbers are introduced as~\cite{Myshkis1987} 
\begin{equation}\label{eq:BoNumbers}
        \text{Bo }= \frac{\rho_L a R_A^2}{\gamma},  \quad
        \text{P} = \frac{\rho_L \omega^2 R_A^3}{2\gamma}, 
\end{equation}
to describe the ratio of inertial to capillary pressures, with $a$ and $\omega$ being the linear acceleration and angular rotation rate associated with different perturbations, respectively.

\subsubsection{Tidal Effects}\label{sec:methods/Perturbations/tidal_Accel}

Due to its spatial extent, the telescope experiences differential gravitational forces across its structure. The motion of a fluid element located at ${}^{\mathcal{G}}\vec{r} = [x_G,y_G,z_G]^T$ relative to the spacecraft center of mass, in the orbital Hill frame $\mathcal{G}$, is governed by the unforced Hill--Clohessy--Wiltshire equations~\cite{Schaub2018,Vallado2013}
\begin{subequations}\label{eq:hcw_full}
\begin{align}
\ddot{x}_G - 2n\dot{y}_G - 3n^2 x_G &= 0, \\
\ddot{y}_G + 2n\dot{x}_G &= 0, \\
\ddot{z}_G + n^2 z_G &= 0,
\end{align}
\end{subequations}
where $n = \sqrt{\mu/a^3}$ denotes the orbital mean motion. For a static interface configuration (${}^{\mathcal{G}}\dot{\vec r}=\vec 0$), the resulting tidal acceleration can be written as
\begin{equation}\label{eq:tidal_tensor}
{}^{\mathcal{G}}\vec a_{\text{t}} = {}^{\mathcal{G}}\ddot{\vec{r}} = T_G {}^{\mathcal{G}}\vec{r}
= n^2
\begin{bmatrix}
3 & 0 & 0\\
0 & 0 & 0\\
0 & 0 & -1 
\end{bmatrix}
{}^{\mathcal{G}}\vec{r} ,
\end{equation}
with the tidal gravity--gradient tensor $T_G$.

Since the interface response is formulated in the telescope-fixed frame $\mathcal{B}_\text{Cart}$, the tidal field must be expressed in this frame as well. Rather than considering all possible orientations, the analysis is restricted to the extreme alignments that bound the tidal loading. While intermediate orientations introduce off-diagonal components in the transformed tidal tensor, these terms represent directional coupling rather than an increase in tidal strength. The magnitude of the resulting acceleration field is bounded by the spectral radius of the source tensor $T_G$, whose largest eigenvalue corresponds to the orbital radial gradient $3n^2$. Consequently, alignment with the principal axes of $T_G$ captures the maximum tidal loading experienced by the interface.

In the worst-case configuration (WC), the telescope’s radial direction aligns with the orbital radial gradient, yielding maximum in-plane loading. In the best-case configuration (BC), this gradient aligns with the optical axis, reducing radial forcing. The corresponding accelerations on a fluid element located at ${}^{\mathcal{B}}\vec{r} = [x_B,y_B,z_B]^T$ are
\begin{subequations}\label{eq:tidal_acceleration}
\begin{align}
{}^{\mathcal{B_\text{Cart}}}\vec a_{\text{t,WC}} &= 3n^2 x_B \,\vec{\hat{b}}_1
- n^2 (z_B-z_{B,\text{CoM}})\,\vec{\hat{b}}_3,\\
{}^{\mathcal{B}_\text{Cart}}\vec a_{\text{t,BC}} &= -n^2 x_B \,\vec{\hat{b}}_1
+ 3n^2 (z_B-z_{B,\text{CoM}})\,\vec{\hat{b}}_3,
\end{align}
\end{subequations}
where $z_{B,\text{CoM}}$ denotes the axial center-of-mass location.

\subsubsection{Solar Radiation Pressure}\label{sec:methods/Perturbations/SRP}

Solar radiation pressure (SRP) arises from the momentum transfer of incident and reflected solar photons and induces a continuous acceleration. For a spacecraft equipped with a sunshield of Sun-facing area $A_s \approx A_{\text{mirror}}$, the resulting force can be written as~\cite{wie2008space,LUMIO}
\begin{equation}
\begin{split}
\vec{F}_{\text{SRP}}
= -\frac{S A_s}{c}(\vec{e}_S \cdot \vec{e}_n)
&\Bigl[(1-\rho_s)\vec{e}_S \\
&+ \left(2\rho_s(\vec{e}_S \cdot \vec{e}_n)
  +\tfrac{2}{3}\rho_d\right)\vec{e}_n
\Bigr],
\end{split}
\end{equation}
where $S$ is the solar irradiance, $c$ the speed of light, $\vec{e}_S$ the Sun-line direction, and $\vec{e}_n$ the outward normal of the sunshield, enclosing an angle $\beta$ such that $\vec{e}_S\!\cdot\!\vec{e}_n=\cos\beta$. The coefficients $\rho_s$ and $\rho_d$ denote the specular and diffuse reflectivity, respectively.

For typical thermal-control surfaces (e.g.\ silvered Teflon or VDA), the absorptance is low  ($\alpha \approx 0.1$) and $\rho_d \ll \rho_s$, yielding $\rho_s \approx 1-\alpha \approx 0.9$~\cite{Gilmore2002}. In the nominal configuration, the sunshield is oriented perpendicular to the Sun-line ($\beta \approx 0$), such that the dominant SRP contribution is aligned with the telescope axis and acts as an effective axial load,
\begin{equation}\label{eq:F_SRP}
    {}^{\mathcal{B}} \vec F_{\text{SRP,ax}} = \frac{S\,A_s\,(1+\rho_s)}{c} \vec{\hat{b}}_3,
\end{equation}
with the corresponding SRP-induced acceleration of the spacecraft of $a_{\text{SRP,ax}} =F_{\text{SRP,ax}} / m_{\text{SC}}$. In the spacecraft frame the liquid therefore experiences an effective body force opposite to the SRP-induced acceleration onto the dish. Thus, the corresponding inertial potential at position ${}^{\mathcal{B}}\vec{r} = [x_B,y_B,z_B]^T$ entering the fluid equations is
\begin{equation}
    \Pi_{\text{SRP}}
    = -\frac{F_{\text{SRP}}\cos^2\!\beta}{m_{\text{SC}}}\,z_B.
\end{equation}

Due to the axial symmetry of the telescope, SRP does not produce a net torque about the optical axis.  However, an offset between the center of pressure of the sunshield and the spacecraft center of mass generates disturbance torques about transverse axes. In practice, these torques must be counteracted by the attitude control system to maintain the nominal pointing configuration and are therefore neglected in the present analysis.

\subsubsection{Self-Gravitation}\label{sec:methods/Perturbations/SelfGrav}
The self-gravitational potential $\Pi_{\text{sg}}$ accounts for mutual attraction between fluid elements and the spacecraft structure. The service bus is approximated as a point mass $m_{\text{SB}}$, while the mirror is modeled as an axisymmetric body of uniform density. The point-mass potential of the service bus at position ${}^{\mathcal{B}}\vec{r} = [x_B,y_B,z_B]^T$ is~\cite{Schaub2018}
\begin{equation}
\Pi_{\text{sg,bus}}
= -\frac{G m_{\text{SB}}}{\sqrt{r_B^2 + (z_B-z_{\text{B,SB}})^2}},
\end{equation}
with $G$ being the gravitational constant.

The potential of the mirror is expressed through a combination of section ($\Omega$) and boundary ($\partial\Omega$) integrals~\cite{2011Trova_SelfGrav}
\begin{equation}\label{eq:self-grav-mirror}
\begin{aligned}
\Pi_{\text{sg,mirror}}
= -2 \rho_L G \Biggl(
\int_{(\partial \Omega)}
T_B \, dR_B - \int_{(\Omega)}
T_S \, dZ_B \, dR_B 
\Biggr),
\end{aligned}
\end{equation}
with
\begin{subequations}\label{eq:self-grav-mirror-terms}
\begin{align}
    T_B &= \sqrt{\frac{R_B}{r_B}} (Z_B - z_B) k \bigl[K(k) - (1 - o^2) M(o,k)\bigr], \\
    T_S &= \sqrt{\frac{R_B}{r_B}} k E(k),
\end{align}
\end{subequations}
and $K(k)$, $E(k)$, and $M(k)$ being the complete elliptic integrals of the first, second, and the third kind, respectively, with the arguments
    \begin{equation}
        k = \sqrt{\frac{4r_BR_B}{(r_B+R_B)^2+(z_B-Z_B)^2}}, \qquad o = \frac{2\sqrt{r_BR_B}}{r_B+R_B}.
    \end{equation}
This expression of the potential of the liquid mirror depends on the position of the liquid, coupling its interface with the distribution of the liquid mass. This interaction, similar to the coupling found for example in highly susceptible ferrofluids~\cite{RomeroCalvo2022_FerrohydrodynamicInterfaceTracking,RomeroCalvo2020MagneticLiquids,RomeroCalvo2021_AxisymmetricFerrofluid}, is addressed with an iterative solution. The total self-gravitational potential is the linear superposition of the two mass components $\Pi_{\text{sg}} = \Pi_{\text{sg,bus}} + \Pi_{\text{sg,mirror}}$. The contribution $\Pi_{\text{sg,mirror}}$ incorporates the self-gravitation imposed by the fluid and the substrate, both determined separately with Eq.~\ref{eq:self-grav-mirror}.

\subsubsection{Electrostatic Effects}\label{sec:methods/Perturbations/Electrostatic_Effects}

Spacecraft exposed to solar winds reach a floating potential $\phi_\text{sc}$ through the balance of ambient plasma, photoelectron, and secondary and backscattered electron currents. Long-term measurements near 1~AU yield nominal potentials of $\phi_\text{sc} \approx 5$--$13$~V, with maxima of 30~V~\cite{Wilson2023}. Without electrical insulation, conductive liquids equilibrate to the potential of the spacecraft, $\phi_\text{sc}$. Approximating the shallow spherical cap ($z_0/R_A \approx 1/8$) as a flat disc, the surface charge at position ${}^{\mathcal{B}}\vec{r} = [x_B,y_B,z_B]^T$ follows the classical conducting-disc distribution~\cite{Smirnov2024, Jackson1999}
\begin{equation}
    \zeta(r_B) = \frac{2\varepsilon_0\,\phi_\text{sc}}{\pi\sqrt{R_A^2 - r_B^2}},
    \label{eq:disc_charge}
\end{equation}
where $\varepsilon_0$ is the vacuum permittivity. Applying Gauss's law and the Maxwell stress tensor yields the electrostatic pressure normal to the free surface~\cite{Landau1984}
\begin{equation}
    P_e(r_B) = \frac{\zeta^2}{2\varepsilon_0}
           = \frac{2\varepsilon_0\,\phi_\text{sc}^2}{\pi^2(R_A^2 - r_B^2)}.
    \label{eq:maxwell_stress}
\end{equation}
Evaluating at $r_B = 0.99\,R_A$ to obtain a representative near-rim estimate while avoiding the classical singularity at $r_B = R_A$, and distributing over the film mass per unit area $\rho_L h$, gives
\begin{equation}
    a_{\text{ES}} = \frac{P_e(0.99\,R_A)}{\rho_L h_0}
        \approx \frac{100\,\varepsilon_0\,\phi_\text{sc}^2}{\pi^2\,\rho_L\,h_0\,R_A^2}.
    \label{eq:a_e}
\end{equation}

\subsubsection{Other Perturbations}\label{sec:methods/Perturbations/Other_Perturbs}

Internal spacecraft disturbances, such as reaction wheel imbalance, mechanism activity, and thermoelastic motion, can induce additional perturbations. For JWST, integrated structural–optical analyses identify the reaction wheel assemblies as the dominant sources of micro-vibration and jitter, exciting both local and global structural modes of the telescope \cite{Hyde2004,Liu2013AIAA_2008_7232}. 

Additional g-jitter disturbances arise from cryocoolers, stepper motors, deployment mechanisms, and other periodic drives. These introduce broadband micro-vibrations that are analyzed using frequency-domain force inputs and structural mode propagation rather than as steady accelerations \cite{flywheel_microvibration_2021,Nelson1994GJitter}. 

Because none of these effects are published in terms of linear accelerations, translating them into equivalent Bond numbers or axisymmetric body forces is not possible without mission-specific data. Nevertheless, JWST heritage demonstrates that internal disturbances constitute a significant design driver. For LMTs, whose free interfaces are more sensitive than rigid mirrors, such disturbances must be addressed in future coupled structural–fluid studies, though they cannot be accurately quantified in the present work.


\subsection{Micrometeoroid Impact Crater Model}
\label{sec:methods/impact}

At orbital impact velocities, the specific kinetic energy of the impacting particle exceeds the vaporization enthalpy of both materials, placing the event in the shock-dominated regime \cite{melosh1989impact}. Crater penetration depth $d_P$ is estimated from impedance-matched Hugoniot shock relations \cite{marsh1980lasl} combined with a hydrodynamic penetration model~\cite{birkhoff1948explosives},
\begin{equation}
    d_P = \frac{u_p}{v_P - u_p}\, D_P,
    \label{eq:craterdepth}
\end{equation}
where $v_P$ is the impact velocity and $D_P$ the impactor diameter. The interface particle velocity $u_p$ is obtained from the Hugoniot pressure-continuity condition at the boundary of impactor and target,
\begin{equation}
    \rho_T\, u_p \left( C_T + S_T\, u_p \right) = \rho_P \left( v_P - u_p \right)\!\left( C_P + S_P \left( v_P - u_p \right) \right),
    \label{eq:impedance}
\end{equation}
by taking the physical root $0 < u_p < v_P$, with material parameters for liquid gallium as a Galinstan surrogate ($C_T = \SI{2740}{m/s}$, $S_T = 1.52$) and basalt for the silicate impactor ($C_P = \SI{3500}{m/s}$, $S_P = 1.50$, $\rho_P = \SI{3000}\,\text{kg}\,\text{m}^{-3}$) \cite{marsh1980lasl}.

The cumulative meteoroid flux at \SI{1}{AU} is given by~\cite{grun1985collisional}
\begin{equation}
    \begin{split}
        F = \bigl(c_4 m^{\gamma_4} + c_5\bigr)^{\gamma_5}
            &+ c_6\bigl(m + c_7 m^{\gamma_6} + c_8 m^{\gamma_7}\bigr)^{\gamma_8} \\
            &+ c_9\bigl(m + c_{10} m^{\gamma_9}\bigr)^{\gamma_{10}},
    \end{split}
    \label{eq:grun}
\end{equation}
where $m$ is the particle mass in grams, $F$ is the cumulative flux in $\text{m}^{-2}\,\text{s}^{-1}$, and with constants $c_4 = 2.2\times10^{3}$, $c_5 = 15$, $c_6 = 1.3\times10^{-9}$, $c_7 = 10^{11}$, $c_8 = 10^{27}$, $c_9 = 1.3\times10^{-16}$, $c_{10} = 10^{6}$, $\gamma_4 = 0.306$, $\gamma_5 = -4.38$, $\gamma_6 = 2$, $\gamma_7 = 4$, $\gamma_8 = -0.36$, $\gamma_9 = 2$, and $\gamma_{10} = -0.85$.

\subsection{Numerical Implementation}
\label{sec:methods/NumericalSolution}

\subsubsection{Interface Parameterization and Boundary Conditions}
\label{sec:methods/InterfaceParameterization&BC}

When limiting the analysis to axisymmetric perturbations, Eq.~\ref{eq:Equilibrium} can be solved in the cylindrical frame $\mathcal{B}$ by adopting an arc-length parameterization of the form $\{r_B(s),z_B(s)\}$, illustrated in Fig.~\ref{fig:flute}. Although the LMT is subject to lateral loads in a real environment, adopting an axisymmetric framework simplifies the problem and enables a first-order feasibility analysis of the concept. Under these conditions, Eq.~\ref{eq:Equilibrium} becomes \cite{Myshkis1987}
\begin{subequations}\label{eq:eqsurf}
    \begin{equation}
        r_B'' = -z_B'\left\{\frac{\rho_L}{\gamma}\left(\Pi+ \Pi_0^*\right)-\frac{z_B'}{r_B}\right\},
    \end{equation}
    \begin{equation}
        z_B'' = r_B'\left\{\frac{\rho_L}{\gamma}\left(\Pi+ \Pi_0^*\right)-\frac{z_B'}{r_B}\right\},
    \end{equation}
where $\Pi_0^*$ is an integration constant, $r_B'$, $z_B'$, $r_B''$, $z_B''$ are first and second order arc parameter derivatives of the interface coordinates $r_B(s)$ and $z_B(s)$. Volume conservation is imposed through    
\begin{equation}
    V'=2\pi r_Br_B'z_B,
\end{equation}
\end{subequations}
with $V(s)$ being the volume below the interface between $s=0$ and $s$. This set of equations creates a second-order boundary-value problem with a free parameter $\Pi_0^*$ and six degrees of freedom. Therefore, six boundary conditions are needed to preserve axisymmetry, volume conservation, and interface pinning, providing
\begin{equation}\label{eq:surf_BCs}
    \begin{aligned}
        s=0: & \quad r_B = 0, \quad r_B' = 1, \quad z_B' = 0, \quad V = 0 \\
        s=s_w: & \quad z_B(r_B) = z_e, \quad V(r_B) = V_0.
    \end{aligned}
\end{equation}
The resulting boundary-value problem is solved numerically using a shooting method similar to those employed in previous studies of axisymmetric ferrofluid deformation and magnetic propellant positioning~\cite{RomeroCalvo2021_AxisymmetricFerrofluid,RomeroCalvo2021_MagneticPositivePositioning}. In this approach, the initial interface height $z_0=z_B(s{=}0)$ and the integration constant $\Pi_0^*$ are adjusted iteratively until the edge-height constraint $z_B(R_A)=z_e$ and the volume constraint $V(R_A)=V_0$ are satisfied. Convergence is declared when the residuals in edge height and enclosed volume fall below $10^{-13}$~m and $10^{-13}$~m$^3$, respectively. During each equilibrium solve, the external force potentials entering $\Pi$ (including the self-gravitational field) are held fixed and updated only between outer iterations.

A suitable initial guess for the integration constant $\Pi_0^*$ may be obtained analytically by looking at an unperturbed system. Without perturbations, the integration constant becomes $\Pi_0^* = {2\gamma}/{(\rho_L R_0)}$ according to Eq.~\ref{eq:Equilibrium}.

\subsubsection{Self Gravitation Potential Evaluation}
\label{sec:methods/SelfGravNumericalImplement}
The fluidic contribution to the mirror self-gravitational potential $\Pi_{\mathrm{sg,mirror}}$ depends on the free-surface shape $z_B(r_B)$, which in turn follows from the capillary balance including that same potential. The problem is therefore nonlinear and is solved iteratively. Starting from the self-gravitation-free equilibrium shape, the potential is evaluated, the interface is updated via the equilibrium solver, and the procedure is repeated until the interface position converges. The complete elliptic integrals in Eq.~\ref{eq:self-grav-mirror} are evaluated using Carlson's symmetric forms~\cite{Carlson1995}, with parameters clamped to $[10^{-12},\,1-10^{-12}]$ to avoid singularities. This clamping has no measurable effect on the converged shape. To avoid repeated quadrature during the shooting-based solve, the potential and its spatial gradients are precomputed on an adaptive $(r,z)$ grid and supplied through interpolation. The discretization of this grid is uniform and fixed in the radial direction and adapted between iterations in the vertical direction by contracting the grid height based on the maximum interface displacement observed between successive iterates, concentrating resolution near the free surface. The potential evaluation is implemented in C++ and interfaced with MATLAB~\cite{MATLAB2023} via a MEX routine with OpenMP parallelization. A sensitivity analysis comparing two discretizations of differing numerical fidelity yields a surface RMS difference of $1.6\times10^{-5}\,\mathrm{m}$ ($0.33\,\%$ of the RMS perturbation-induced deformation) with comparable convergence. The iteration is advanced until the step size falls below $1\,\%$ of the RMS deformation induced by the perturbation. Further refinement produces no qualitative changes, confirming that the equilibrium surface RMS is effectively converged.

\subsubsection{Numerical Time Integration}
\label{sec:methods/TimeIntegration}

The time-dependent evolution of the liquid interface is obtained by numerically integrating Eq.~\ref{eq:thinfilm_pde} in the spherical frame $\mathcal{S}$. As in the equilibrium analysis, the dynamical simulations are restricted to axisymmetric perturbations to reduce computational complexity and focus on the dominant settling dynamics. Under this assumption ($\partial_\phi = 0$), the surface gradient and divergence operators reduce to their meridional components, and the thin-film equation becomes
\begin{equation}
\frac{\partial h}{\partial t}
=
\frac{1}{R_0^2\sin\theta}
\frac{\partial}{\partial\theta}
\left[
\sin\theta
\frac{h^3}{3\mu}
\frac{\partial}{\partial\theta}
\left(
-2\gamma\mathcal{H}
+\rho\Pi
\right)
\right].
\label{eq:thinfilm_axisym_sph}
\end{equation}
The mean curvature of the axisymmetric interface is computed from the surface geometry as~\cite{Kang2016}
\begin{equation}
2\mathcal{H}
=
\frac{2}{R_0}
+
\frac{1}{R_0^2}
\left[
2h
+
\frac{1}{\sin\theta}
\frac{\partial}{\partial\theta}
\!\left(
\sin\theta\,\frac{\partial h}{\partial\theta}
\right)
\right].
\label{eq:curvature_linearized_sph}
\end{equation}
The spatial domain $\theta \in [0,\theta_C]$ is discretized using a finite-difference scheme on a uniform angular grid. Second-order central differences are employed for all spatial derivatives, with one-sided stencils applied at the boundaries where required. At the axis ($\theta=0$), symmetry conditions enforce $\partial_\theta h = 0$, while at the outer boundary ($\theta=\theta_C$) the interface is pinned to the container wall, consistent with the equilibrium analysis.

The method-of-lines reduces Eq.~\ref{eq:thinfilm_axisym_sph} to a system of time-dependent nonlinear ODEs~\cite{Schiesser1991}, integrated directly using the \texttt{ode15s} solver in MATLAB~\cite{MATLAB2023}. The grid resolution varies across analyses: $N = 5000$ uniformly spaced points in $\theta$ suffice for the thrust-response analysis, whereas $N = 20000$ are required for the post-filling settling analysis, where the shorter spatial wavelength of the perturbation demands a finer resolution of 80 points per sinusoid in the worst case ($R_A = 25$~m, $\lambda_\text{filling} = 0.1$~m). Time integration is performed with absolute and relative tolerances of $10^{-8}$ and $10^{-6}$, respectively.

\subsection{Optical characterization}
\label{sec:methods/Optics_of_LMTs}

Provided the surface is sufficiently reflective at the wavelength of interest, the fluid nature of the substrate is irrelevant to wavefront formation from the standpoint of geometrical optics. The quality of the reflected wavefront is determined strictly by the deviation of the actual liquid surface from its nominal design shape~\cite{BornWolf1999,Cheng2009}. A local surface height error $\delta h$ introduces an Optical Path Difference (OPD) in the reflected wavefront. For paraxial incidence, applicable to high-focal-number systems, the OPD is approximately twice the surface error~\cite{Cheng2009}
\begin{equation}
\mathrm{OPD} \approx 2\,\delta h,
\end{equation}
since the optical path is affected both on incidence and reflection. The standard criterion for high-quality imaging is the Mar\'echal approximation, which relates the Strehl ratio $S$ to the RMS wavefront error $\sigma_{\mathrm{WFE}} = \sqrt{\langle \mathrm{OPD}^2 \rangle}$~\cite{Cheng2009}
\begin{equation}
    \mathcal{S} \approx \exp\!\left[-\left(\frac{2\pi\sigma_{\mathrm{WFE}}}{\lambda}\right)^2\right],
\end{equation}
with $\lambda$ denoting the wavelength of interest. The Strehl ratio describes the ratio of the peak intensity of the aberrated image to that of an ideal, aberration-free counterpart. While the Mar\'echal criterion defines a Strehl ratio of 0.80 as the rigorous diffraction limit~\cite{Mahajan1983}, practical designs often adopt slightly relaxed tolerances to account for environmental factors~\cite{Cheng2009}. In this work, a relative image intensity of $S \approx 0.67$ is considered acceptable~\cite{Cheng2009}. Solving the Mar\'echal equation and converting to RMS surface error $\sigma_h = \sigma_{\mathrm{WFE}}/2$ yields a surface-error budget of
\begin{equation}
\sigma_h \le \frac{\lambda}{20}.
\end{equation}
Introduction of a secondary optic could, in principle, correct known static aberrations and partially address the error budget, though this lies beyond the scope of the present work.


\section{Results \& Discussion}\label{sec:resdis}

Telescopes face strict thermal constraints to observe faint astrophysical sources against the zodiacal background~\cite{rigby2023}. The mirror surface, as a greybody emitter, must not contribute thermal photons that overwhelm this natural background. Similarly, detectors must be cooled below a material-dependent threshold set by their band-gap energy so that thermally generated dark current remains negligible compared to the photon signal~\cite{rogalski2005}. For HgCdTe detectors with tuneable cutoff wavelength $\lambda_c$, the required operating temperature scales as $T_{\mathrm{det}} \propto 1/\lambda_c$~\cite{rogalski2005}. Fixed-gap materials (Si:As, Ge:Ga) demand progressively lower temperatures as their spectral response extends further into the infrared~\cite{ressler2015,poglitsch2010}.

\begin{figure}[t!]
\centering
\includegraphics[width=\linewidth]{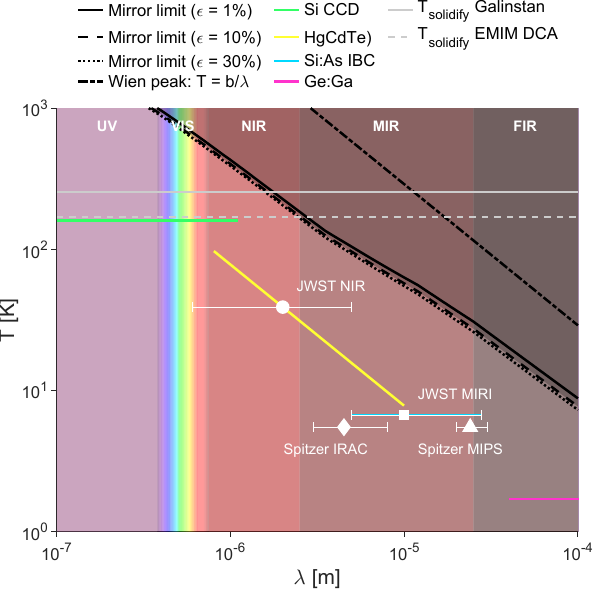}
\caption{Maximum permissible mirror temperature as a function of observing wavelength, derived from Eq.~\ref{eq:Tmirror} with $f = 1\%$ and three representative emissivities. Colored lines indicate detector operating temperatures for the indicated material families~\cite{rogalski2005}. Operating points are shown for JWST~\cite{wright2015}, Spitzer~\cite{werner2004}, and Herschel~\cite{pilbratt2010}.}
\label{fig:TemperatureRequirements}
\end{figure}

The zodiacal spectrum is modeled using tabulated intensities from Leinert et al.~\cite{leinert1998} at solar elongation $90^\circ$ in the ecliptic, scaled to the ecliptic pole as a worst case scenario. Tabulated values are interpolated in log--log space. The maximum permissible mirror temperature is obtained by requiring that the spectral radiance of the mirror remains below a fraction $f$ of this background and inverting the Planck function~\cite{rieke2003}
\begin{equation}
T_{\mathrm{mirror,max}}(\lambda) = \frac{hc}{k\lambda}\left[\ln\!\left(1 + \frac{\varepsilon}{f}\,\frac{2hc^2/\lambda^5}{I_{\mathrm{zodi}}(\lambda)}\right)\right]^{-1},
\label{eq:Tmirror}
\end{equation}
where $h$ is the Planck constant, $k$ the Boltzmann constant, and $I_{\mathrm{zodi}}$ the tabulated zodiacal background spectral radiance~\cite{leinert1998}. Figure~\ref{fig:TemperatureRequirements} maps the maximum mirror temperature over the wavelength for representative emissivities and $f = 1\%$. For Galinstan ($\varepsilon = 30\%$, $f = 1\%$), the maximum observable wavelength based on its solidification temperature of 254~K~\cite{HANDSCHUHWANG2022101642} is $\lambda_\text{max,G} = 1.59~\mu$m. By switching to an ionic liquid such as 1-ethyl-3-methylimidazolium dicyanamide (EMIM DCA) (glass temperature of 169~K~\cite{MacFarlane2002}) with a silver reflective layer~\cite{Rowlands2024_SelfAssemblingILM} ($\varepsilon = 1\%$), this limit extends to $\lambda_\text{max,IL} = 2.75~\mu$m, covering the entire visible and near-infrared spectrum. The results presented hereafter are scoped to this range.

\subsection{Equilibrium Interface}\label{sec:resdis/Static_Response}

\begin{table}[b!]
\centering
\caption{Maximum perturbation-induced accelerations and Bond numbers (Eq.~\ref{eq:BoNumbers}) for tidal effects in SEL2, electrostatic effects, solar radiation pressure ($\beta=0$°), and self-gravitation (broken down by fluid, dish, and bus) for selected apertures.}

\begin{tabular}{lccc c}
\toprule
{Parameter} & {$D_A=5$ m} & {$D_A=15$ m} & {$D_A=50$ m} &
{Units}\\
\midrule

$a_{\text{SRP}}$
& $2.4\!\times10^{-8}$ & $7.9\!\times10^{-8}$ & $11\!\times10^{-8}$
& $[\text{m/s}^2]$ \\
$\mathrm{Bo}_{\text{SRP}}$
& $1.3\!\times\!10^{-3}$ & $0.040$ & $0.60$
& $[\,\;\;-\;\;\,]$ \\
\addlinespace[4pt]
$a_{\text{sg, fluid}}$
& $5.5\!\times10^{-8}$ & $6.5\!\times10^{-8}$ & $7.1\!\times10^{-8}$
& $[\text{m/s}^2]$ \\
$\mathrm{Bo}_{\text{sg, fluid}}$
& $3.1\!\times\!10^{-3}$ & $0.032$ & $0.40$
& $[\,\;\;-\;\;\,]$ \\
\addlinespace[4pt]
$a_{\text{sg, dish}}$
& $1.0\!\times10^{-8}$ & $1.2\!\times10^{-8}$ & $1.5\!\times10^{-8}$
& $[\text{m/s}^2]$ \\
$\mathrm{Bo}_{\text{sg, dish}}$
& $5.7\!\times\!10^{-4}$ & $6.2\!\times\!10^{-3}$ & $0.085$
& $[\,\;\;-\;\;\,]$ \\
\addlinespace[4pt]
$a_{\text{sg, bus}}$
& $2.3\!\times10^{-8}$ & $2.3\!\times10^{-8}$ & $2.3\!\times10^{-8}$
& $[\text{m/s}^2]$ \\
$\mathrm{Bo}_{\text{sg, bus}}$
& $1.3\!\times\!10^{-3}$ & $0.011$ & $0.13$
& $[\,\;\;-\;\;\,]$ \\
\addlinespace[4pt]
$a_{\text{ES, 30V}}$
& $203\!\times10^{-12}$ & $22.0\!\times10^{-12}$ & $2.0\!\times10^{-12}$
& $[\text{m/s}^2]$ \\
$\mathrm{Bo}_{\text{ES, 30V}}$
& $1.1 \times 10^{-5}$ & $1.1 \times 10^{-5}$ & $1.1 \times 10^{-5}$
& $[\,\;\;-\;\;\,]$ \\
\addlinespace[4pt]
$a_{\text{tidal}}$
& $3.3\!\times10^{-13}$ & $9.0\!\times10^{-13}$ & $30\!\times10^{-13}$
& $[\text{m/s}^2]$ \\
$\mathrm{Bo}_{\text{tidal}}$
& $1.8\!\times\!10^{-8}$ & $4.5\!\times\!10^{-7}$ & $1.7\!\times\!10^{-5}$
& $[\,\;\;-\;\;\,]$ \\
\bottomrule
\end{tabular}
\label{tab:BoAccelerationComparison}
\end{table}

The equilibrium interface shape of a capillary-driven LMT results from the interplay of surface tension and the quasi-static perturbations introduced in Sec.~\ref{sec:methods/Perturbations}. In order to reach optical compliance, this equilibrium shape has to match the target figure up to the surface error criterion defined in Sec.~\ref{sec:methods/Optics_of_LMTs}. Since equilibrium interfaces of fluids in microgravity tend to adopt quasi-spherical shapes, a spherical mirror is considered as the target profile for capillary-driven LMTs.

Assessing the Bond numbers (Eq.~\ref{eq:BoNumbers}) associated with each perturbation source provides a first-order measure of their relative importance and identifies the dominant contributions in the Sun-Earth L2 environment. Table~\ref{tab:BoAccelerationComparison} presents maximum acceleration levels as a function of aperture for different perturbations and their respective Bond numbers. In this calculation, the worst-case spacecraft orientation is considered for tidal effects, electrostatic effects are assessed at an aperture radius of $r = 0.99\,R_A$,  and solar radiation pressure is assessed for normal solar incidence ($\beta = 0$°). Accelerations due to both tidal effects and electrostatic charge are orders of magnitude below those due to SRP and self-gravitation across all apertures assessed. Therefore, tidal and electrostatic effects are omitted in the subsequent discussion.

\begin{figure}[t!]
    \centering
    \includegraphics[width=\linewidth]{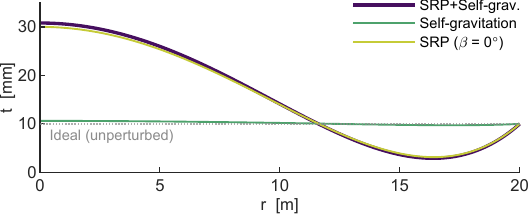}
    \caption{Equilibrium liquid film thickness over the radius of a 40~m aperture Galinstan-based LMT, for solar radiation pressure ($\beta = 0$°) and self-gravitation, both individually and in combination.}
    \label{fig:LiquidInterfaces}
\end{figure}

While the accelerations due to SRP and self-gravitation appear small at magnitudes of tens of nm\,s$^{-2}$, the associated Bond numbers result in significant deviations from the spherical target. Fig.~\ref{fig:LiquidInterfaces} confirms this initial observation. Equilibrium interfaces are shown for SRP ($\beta = 0$°) and self-gravitation individually and combined for a 40~m aperture LMT. In this configuration, surface deformations from the spherical baseline are of the order of $\mathcal{O}(10^{-2})$~m, well beyond acceptable optical limits.

\begin{figure}[b!]
    \centering
    \includegraphics[width=\linewidth]{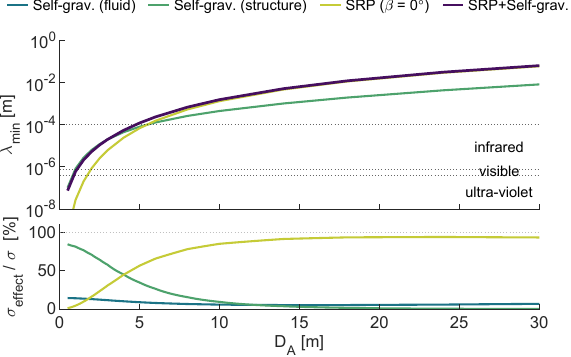}
    \caption{Shortest observable wavelength as a function of aperture for a Galinstan-based liquid mirror with film thickness $h_0=10$~mm, based on the maximum surface deviation under all considered accelerations at SEL2 and a $\lambda/20$ surface error criterion.}
    \label{fig:LambdaVsAperture}
\end{figure}

To determine the aperture range over which the $\lambda/20$ criterion is satisfied, the equilibrium interface is derived for increasing apertures, and the resulting RMS deviation from the optimal spherical surface $\sigma_h$ is extracted. This process yields wavelength-dependent aperture bounds, shown in Fig.~\ref{fig:LambdaVsAperture} for a film thickness of $h_0 = 10$~mm. The absolute deformation increases rapidly with aperture size and exceeds the $\lambda/20$ criterion for visible light ($\lambda=380$~nm) at apertures of 0.86~m. Infrared ($\lambda = 1.5$~um) and far-infrared observations ($\lambda = 2.5$~um) remain feasible up to apertures of 1.34~m and 1.58~m, respectively. These aperture limits could, in principle, be increased by introducing a secondary optic to correct known static aberrations.

At apertures of approximately 4 m or less, structural self-gravitation (fluid dish and spacecraft bus) is the dominant deformation source. Solar radiation pressure becomes dominant at larger apertures. Neglecting the computationally expensive, iteratively resolved self-gravitational potential of the liquid leads to an underestimate of surface deformation of up to 15\% at $R_A < 2$~m and less than 10\% at larger apertures. This effect is less significant for lower-density liquids and thinner films. Given the large computational overhead and diminishing impact of the self-gravitational potential of the fluid, its effect is omitted in all subsequent parameter sweeps. The results presented are therefore conservative, as they underestimate the true deformation. At lower surface tensions or larger apertures, where interface deformations grow, the self-gravitational acceleration of the fluid can become significant and may ultimately exceed the solar radiation pressure contribution. This is driven by a positive feedback: pooling of liquid at the mirror center increases the local gravitational pull, which in turn promotes further pooling.
\begin{figure}[t!]
    \centering
    \includegraphics[width=\linewidth]{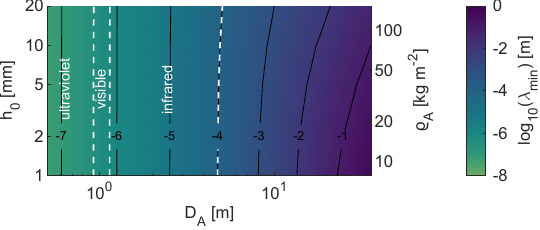}
    \caption{Shortest observable wavelength as a function of aperture and film thickness for a Galinstan-based liquid mirror, based on a $\lambda/20$ maximum surface deviation under SRP ($\beta=0$°) and self-gravitational (fluid dish and spacecraft bus) perturbations.}
    \label{fig:LambdaVsApertureAndFilmThickness}
\end{figure}

The influence of film thickness is shown in Fig.~\ref{fig:LambdaVsApertureAndFilmThickness}, where SRP and self-gravitational effects induced by the spacecraft structure (bus + dish) are considered, and the shortest observable wavelength is plotted as a function of aperture and film thickness. At small apertures, film thickness has a negligible effect, consistent with structural self-gravitation being the dominant perturbation source. As aperture increases, a thicker film yields access to shorter observable wavelengths, since the associated increase in spacecraft mass reduces the SRP-induced acceleration ($a_{\text{SRP}} \propto m_{\text{SC}}^{-1}$). This benefit, however, comes at the cost of increased telescope mass: the areal density of the fluid layer, $\varrho = m_{\text{mirror}}\,A_{\text{mirror}}^{-1}$, increases linearly with film thickness and surpasses the effective value of JWST ($\varrho_{\text{JWST}} = 28.2$~kg\,m$^{-2}$)\footref{fn:nasa} at a thickness of 3.61~mm.

The spacecraft bus model is sized based on JWST and becomes increasingly inaccurate at small apertures, where it overestimates bus mass and thus underestimates the SRP acceleration. Therefore, results below $R_A \approx 1.5$~m should be interpreted with caution. Since the primary motivation for liquid mirror telescopes lies in large-aperture observatories, small apertures are not the regime of interest, but are included to provide a bounding estimate for visible observations.

\subsection{Hydrostatic Stability}
\label{sec:resdis/Stability}

A second constraint on the operational acceleration range is the hydrostatic stability limit, commonly expressed through a critical Bond number $\mathrm{Bo}^*$~\cite{Myshkis1987}. Approximating the liquid mirror as a pinned interface in a cylindrical container under axial acceleration and substituting the rim angle $\theta_C = \arcsin(R_A/R)$ of the unperturbed equilibrium interface as the characteristic angle yields
\begin{equation}
\mathrm{Bo}^* = -0.84 - 2.55\sin(\theta_C) = -1.478.
\end{equation}
Because the liquid mirror interface is pinned at a sharp rim rather than exhibiting a classical contact-angle boundary condition at the wall, this relation can only provide an order-of-magnitude estimate of the critical Bond number and corresponding acceleration levels. However, comparison with the optical surface tolerance limits shows that surface figure errors exceed acceptable levels well before the hydrostatic instability threshold is reached.

The same holds for retention at the rim: pinning edges admit a range of interface slope~\cite{Langbein2002} and tolerate interface height variations of the order of 0.1 to 1~mm for a representative rim geometry, three to five orders of magnitude above the deviations admitted by the $\lambda/20$ criterion. Consequently, neither hydrostatic detachment nor loss of pinning governs the operational acceleration envelope for any configuration that meets the optical requirement across the full aperture.

\subsection{Dynamic Response}
\label{sec:resdis/dynamicresponse}

Beyond the equilibrium interface shape, the dynamic response of the liquid mirror to transient events is of relevance across all mission phases. During initial mirror formation, particle impacts, and propulsive maneuvers, time-dependent deformations are induced in the liquid film by transient forcing. The characteristic timescales of excitation and subsequent settling govern whether the mirror can maintain or recover optical figure within operational limits. The following sections address the dynamic response of the mirror to these events.

\begin{figure}[b!]
    \centering
    \includegraphics[width=\linewidth]{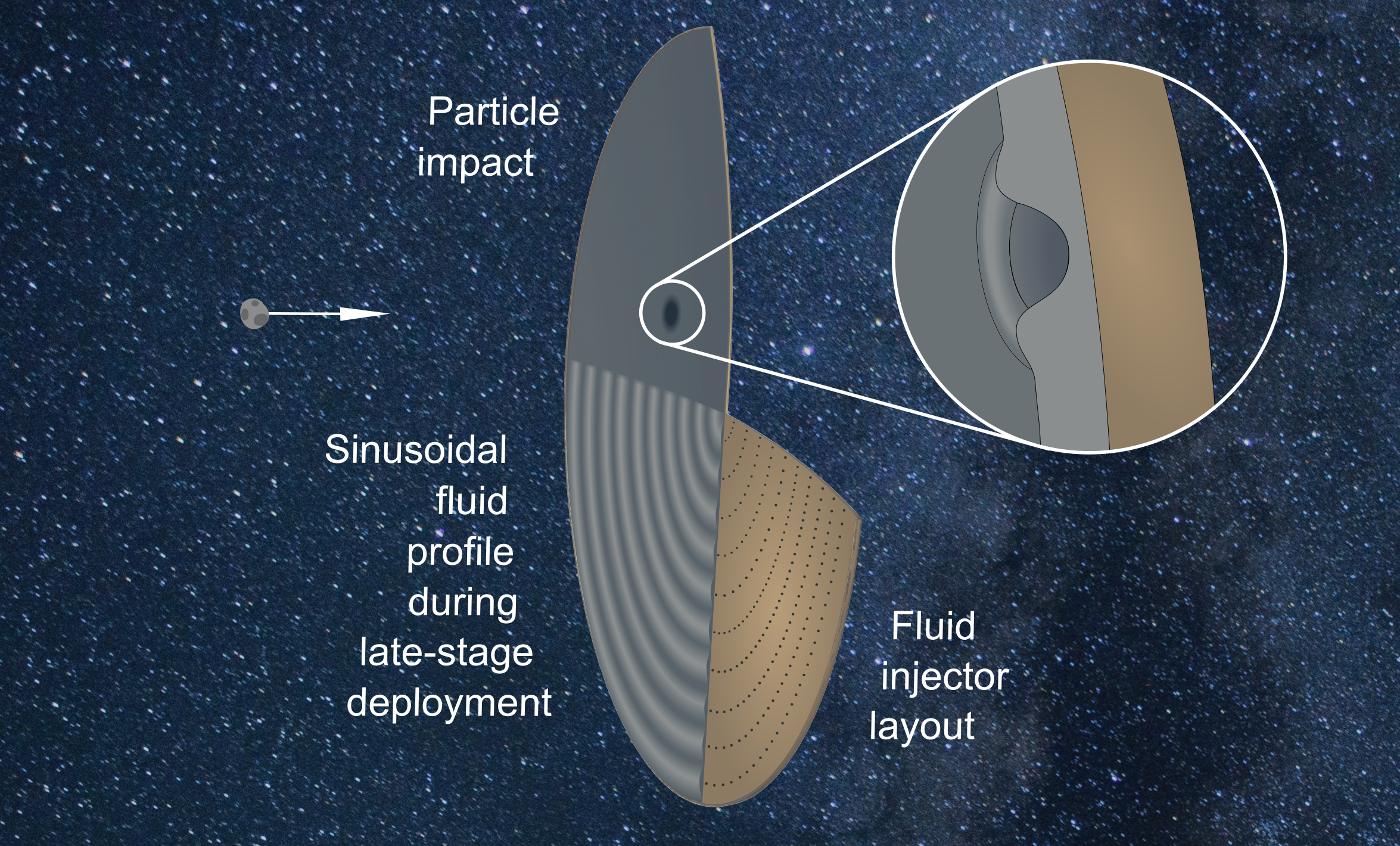}
    \caption{Depiction of assessed dynamic events. }
    \label{fig:DynamicsDepiction}
\end{figure}

\subsubsection{Settling Behavior after Deployment}

\begin{figure}[t!]
    \begin{subfigure}[t]{0.57\linewidth}
        \centering
        \includegraphics[width=\linewidth]{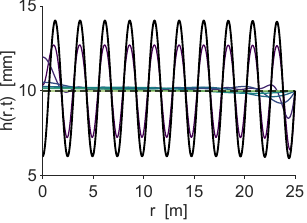}
        \subcaption{Film thickness profiles}
    \end{subfigure}
    \hfill
    \begin{subfigure}[t]{0.42\linewidth}
        \centering
        \includegraphics[width=\linewidth]{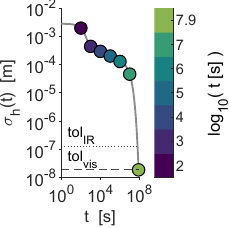}
        \subcaption{RMS surface error}
    \end{subfigure}
    \caption{Capillary relaxation of a sinusoidal surface perturbation on a 50~m aperture Galinstan-based mirror ($h_0 = 10\,\text{mm}$, 4~mm amplitude, 10.25 cycles). (a) depicts film thickness profiles at the times indicated in (b) by the colored markers, with the initial condition shown in black and the equilibrium thickness $h_0$ as a dashed line; (b) shows the RMS surface deviation $\sigma_h(t)$ relative to the undisturbed film, with the dashed line marking the $\lambda_\text{vis}/20$ tolerance.}
    \label{fig:SingleSinus_relaxation}
\end{figure}

To estimate the relaxation of residual deformations after deployment, the interface is initialized with a sinusoidal thickness modulation and allowed to relax under nominal microgravity conditions. This fluid profile resembles the corresponding injector layout illustrated in Fig.~\ref{fig:DynamicsDepiction}. Figure~\ref{fig:SingleSinus_relaxation} describes a representative settling evolution for a 50~m mirror. Because the model is axisymmetric, discrete injection points are averaged over an annulus. Hence, the resulting settling times are order-of-magnitude estimates of the late-stage redistribution timescale rather than predictions of a resolved deployment sequence.

For the case shown in Fig.~\ref{fig:SingleSinus_relaxation}, the relaxation time to a residual surface error of visible tolerances (380~nm) is of the order of 2.5 years for a 50~m mirror, settling to near-infrared tolerances (2.5~$\mu$m) takes about two years.
\begin{figure}[t!]
    \centering
    \begin{subfigure}{\linewidth}
        \centering
        \includegraphics[width=\linewidth]{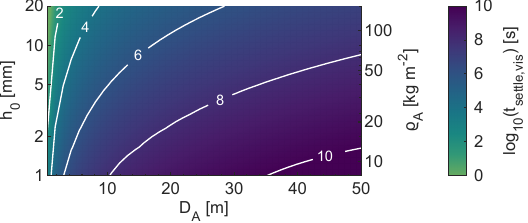}
        \caption{Settling time to the optical RMS tolerance for $\lambda = 380$~nm}
        \label{fig:SinusSettling_Vis}
    \end{subfigure}
    \vspace{0.5cm}
    \begin{subfigure}{\linewidth}
        \centering
        \includegraphics[width=\linewidth]{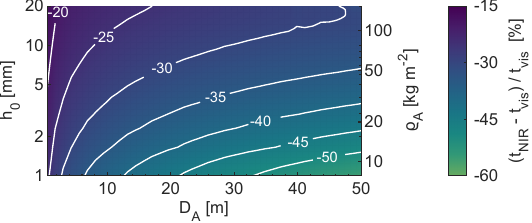}
        \caption{Percentage decrease in settling time for $\lambda = 2.5~\mu$m vs $\lambda = 380$~nm}
        \label{fig:SinusSettling_IR}
    \end{subfigure}
    \caption{Sinusoidal perturbation settling time for a Galinstan-based mirror as a function of film thickness $h_0$ and aperture $D_A$, under zero-gravity pure capillary relaxation.
    The perturbation has a spatial wavelength $\lambda_\text{Filling}=0.1$~m and an amplitude of $0.4\,h_0$. In (a) results are shown for visible tolerances and in (b) the percentage decrease in settling time for near-infrared compared to visible wavelenghts.}
    \label{fig:Sine_Settling}
\end{figure}
The log-log evolution of $\sigma_h$ exhibits three distinct phases, reflecting the successive damping of perturbation components from short to long spatial scales. Initially, the interior mid-aperture peaks are rapidly damped, producing a steep power-law decay. This is followed by a significantly slower phase governed by boundary-dominated deformations at the pinned edge and the mirror center, which is visible as a reduced slope on the log-log scale. 

Once these boundary peaks have largely subsided, the remaining long-wavelength deformations relax with the steepest log-log slope among the three phases, ultimately bringing the surface within optical tolerance. The settling scales with the viscous--capillary ratio
\begin{equation}
\label{eq:scaling-settling}
\tau \sim \frac{\mu}{\gamma},
\end{equation}
which balances viscous resistance against capillary driving. The low dynamic viscosity and high surface tension of Galinstan make it stand out among candidate mirror liquids. Materials with higher $\mu/\gamma$ exhibit qualitatively identical dynamics but with settling times scaled proportionally upward. For an ionic liquid, a close alternative candidate for liquid mirrors ($\gamma \approx 50~\mathrm{mN~m}^{-1}$, $\mu \approx 100~\mathrm{mPa~s}$~\cite{Gomez2006EMIMES}), the predicted settling time increases by a factor of roughly 600 to more than 1500 years to visible tolerances. Thin-film simulations with adjusted material parameters confirm this scaling.

\begin{figure}[t!]
    \centering
    \begin{subfigure}{\linewidth}
        \centering
        \includegraphics[width=\linewidth]{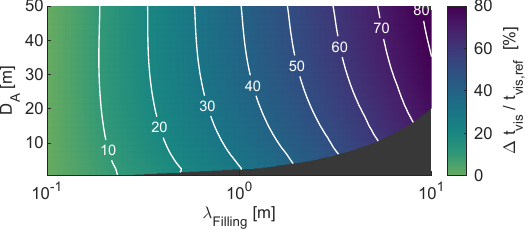}
        \caption{Percentage increase in settling time for increasing spatial wavelength for $\lambda = 380$~nm}
        \label{fig:SinusSettling_Vis_spatialWavelength}
    \end{subfigure}
    \vspace{0.5cm}
    \begin{subfigure}{\linewidth}
        \centering
        \includegraphics[width=\linewidth]{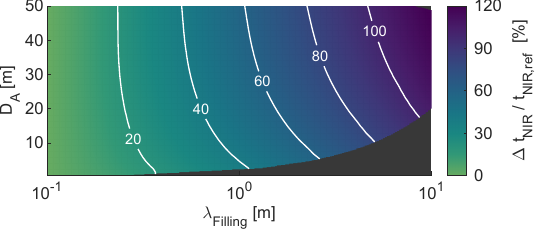}
        \caption{Percentage increase in settling time for increasing spatial wavelength for $\lambda = 2.5~\mu$m}
        \label{fig:SinusSettling_IR_spatialWavelength}
    \end{subfigure}
    \caption{Percentage increase in settling time for sinusoidal perturbations for a Galinstan-based mirror as a function of spatial filling wavelength $\lambda_\text{Filling}$ and aperture $D_A$ at a film thickness $h_0$ of 10~mm, under zero-gravity pure capillary relaxation. Setttling for a spatial wavelength of 0.1~m is used as reference. In (a) results are shown for visible tolerances and in (b) for near-infrared compared to visible. Dark regions mark the limits in spatial wavelength for given apertures.}
    \label{fig:Sine_Settling_Freq}
\end{figure}

The influence of aperture and film thickness on settling time is shown in Fig.~\ref{fig:Sine_Settling}. Thicker films yield shorter settling times, though at the cost of increased mirror mass. Constraining the effective areal density to that of JWST ($\varrho_{\text{JWST}} = 28.2$~kg\,m$^{-2}$)\footref{fn:nasa}, corresponding to a film thickness of 3.61~mm (see Sec.~\ref{sec:resdis/Static_Response}), the settling time crosses the one-year, one-month, and one-day marks at apertures of 23.4/20.9~m, 12.0/10.9~m, and 4.8/4.4~m (NIR/visible), respectively. Figure~\ref{fig:Sine_Settling_Freq} shows the effect of the number of injection points along the arc, quantized by the spatial wavelength $\lambda_\text{Filling}$. Increasing the spatial wavelength (decreasing the number of injection points) leads to an increment in settling time. The settling time under pure capillary relaxation therefore imposes a practical aperture limit on capillary-driven LMTs during initial assembly, with the maximum achievable areal-density governing this limit. Relaxing the limit requires active or passive surface-shaping during deployment.

\subsubsection{Response to Propulsive Maneuvers}
\label{sec:resdis/response}

\begin{figure}[t!]
    \centering
    \includegraphics[width=\linewidth]{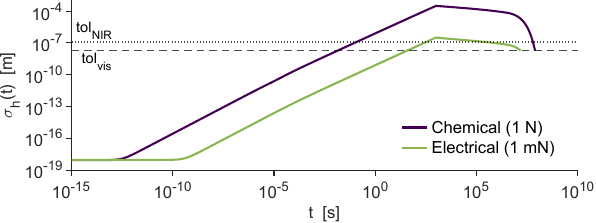}
    \caption{RMS surface error exhibited by the liquid mirror interface when exposed to a $10^{3}$~s uniform axial acceleration from a chemical thruster (1~N) and an electrical thruster (1~mN) for a Galinstan-based 50~m aperture mirror with film thickness $h_0=3$~mm.}
    \label{fig:InterfaceResponseAxial}
\end{figure}

The viscous-capillary dynamics governing interface settling also determine the sensitivity of the mirror to externally imposed accelerations. Two propulsion regimes that bracket the practically relevant design space are considered: chemical reaction-control thrusters at thrust levels of $\mathcal{O}(1)$~N, representative of monopropellant engines used for attitude control and station-keeping (e.g., the MRE-1 class flown on JWST~\cite{Gardner2023_JWST}), and precision low-thrust field-emission electric propulsion (FEEP) at thrust levels of $\mathcal{O}(1)$~mN, representative of fine station-keeping systems~\cite{Krejci2018_FEEP}. 

Figure~\ref{fig:InterfaceResponseAxial} shows the interface response of a 50~m mirror to both regimes. The RMS surface error exceeds the $\lambda/20$ optical tolerance within seconds of load onset, with the stronger forcing shifting this threshold to even shorter timeframes while following the same viscous--capillary growth behavior. After load removal, the interface relaxes on viscous--capillary timescales set primarily by the amplitude of the deformation.

\begin{figure}[t!]
    \centering

    \begin{subfigure}[t]{\linewidth}
        \centering
        \includegraphics[width=\linewidth]{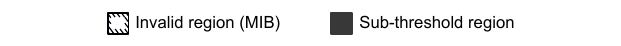}
    \end{subfigure}

    \vspace{0.5em}

    \begin{subfigure}[t]{0.49\linewidth}
        \centering
        \includegraphics[width=\linewidth]{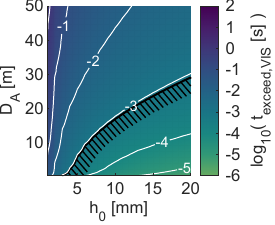}
        \subcaption{Chemical, visible}
    \end{subfigure}
    \hfill
    \begin{subfigure}[t]{0.49\linewidth}
        \centering
        \includegraphics[width=\linewidth]{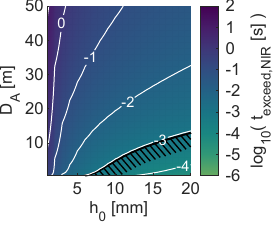}
        \subcaption{Chemical, NIR}
    \end{subfigure}

    \begin{subfigure}[t]{0.49\linewidth}
        \centering
        \includegraphics[width=\linewidth]{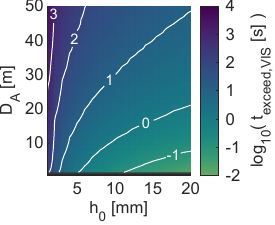}
        \subcaption{Electric, visible}
    \end{subfigure}
    \hfill
    \begin{subfigure}[t]{0.49\linewidth}
        \centering
        \includegraphics[width=\linewidth]{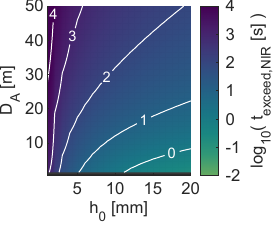}
        \subcaption{Electric, NIR}
    \end{subfigure}

    \caption{Time required to exceed optical limits following a chemical (1 N, top) and electrical (1 mN, bottom) thruster firing as a function of mirror aperture and film thickness, for visible (left) and near-infrared (right) wavelength tolerances.}
    \label{fig:AxialStepResponse_ChemicalThruster}
\end{figure}

Figure~\ref{fig:AxialStepResponse_ChemicalThruster} extends this analysis across aperture and film thickness. Under low-thrust forcing, the RMS surface error remains below the optical tolerance for up to several hours in the near-infrared and several minutes in the visible. Under chemical-thruster forcing, optical limits are exceeded on second to sub-second timescales. This trend opposes that of passive settling: larger apertures and lower film thicknesses are beneficial, yielding longer times before the optical figure is compromised. For the chemical thruster the minimum impulse bit (MIB) $\Delta I_\text{MIB} = F_{\text{Thruster}} \, \Delta t_{\text{Thruster,min}}$ providable by the thrusters is considered, rendering regions where surface errors rise to optical limits below 1~ms infeasible.

In addition to translational accelerations, slewing maneuvers introduce centrifugal in-plane forcing. \cite{Gabay2025_FluidDynamicsLiquidMirror} found that the deformation grows as $t^{3/4}$ during actuation and decays as $t^{-1/4}$ during relaxation, propagating inward from the rim rather than affecting the full aperture uniformly. For their ionic liquid candidate ($\gamma \approx 50~\mathrm{mN\,m^{-1}}$, $\mu \approx 100~\mathrm{mPa\,s}$) and a 50~m mirror operated over ten years with daily slewing maneuvers, the 80\% inner aperture remains below the 20~nm optical threshold for more than 20~years even as the outer rim accumulates deformations of several micrometers. For Galinstan, the lower kinematic viscosity reduces the settling timescale by approximately three orders of magnitude (see Eq.~\ref{eq:scaling-settling}), shrinking the envelope to 7.3 days.

These results reinforce a fundamental trade-off in liquid selection: low viscosity accelerates both initial settling and post-maneuver recovery but also increases sensitivity to operational disturbances, whereas higher viscosity attenuates transient perturbations at the cost of substantially longer passive settling times. Active surface control during maneuvers could relax this trade-off by suppressing deformation growth independently of the bulk liquid properties.

\subsubsection{Settling Behavior after Particle Impacts}
\label{sec:resdis/settling}

\begin{figure}[t!]
    \centering

    \begin{subfigure}[t]{\linewidth}
        \centering
        \includegraphics[width=\linewidth]{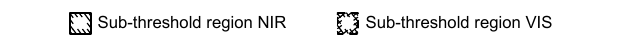}
    \end{subfigure}

    \vspace{0.5em}

    \begin{subfigure}[t]{\linewidth}
        \centering
        \includegraphics[width=\linewidth]{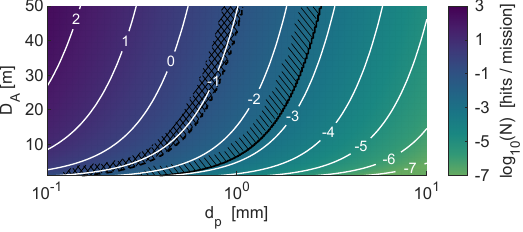}
    \end{subfigure}
    \caption{Expected number of micrometeoroid impact events per mission duration as a function of crater depth $d_P$ and aperture diameter $D_A$. Impactors are modeled as asteroidal silicate particles with a log-normal velocity distribution ($\bar{v} = \SI{17}{km/s}$, $\sigma_{\ln v} = 0.45$). Regions were RMS surface error remains below optical limits for visible light (380~nm) and near-infrared (2.5~um) are marked.}
    \label{fig:ImpactCraterSizeDistribution}
\end{figure}

Due to the high relative velocity of particles impacting the liquid mirror, energy is not transferred into the bulk by mechanical momentum transfer but rather by shock-imposed phase transition of the material. As the specific kinetic energy of the impact exceeds the specific vaporization enthalpy of both impactor and target, both are vaporized into a small plasma bubble that expands and disperses, leaving a roughly hemispherical surface crater~\cite{Fletcher2015,Hill1995HypervelocityAngle}. At hypervelocity, target materials respond hydrodynamically regardless of phase~\cite{melosh1989impact}, making solid-target crater morphology a valid analogue for liquid surfaces in this regime~\cite{crawford1993hypervelocity}. By converting the cumulative flux $F(m)$ (Eq.~\ref{eq:grun}) to a differential flux in impactor diameter space $d_P$ via $f(d_P) = \lvert\mathrm{d}F/\mathrm{d}m\rvert \cdot (\pi/2)\,\rho_P d_P^2$, assuming spherical impactors of density $\rho_P$, and numerically integrating $f(d_P)\,\phi(v)$ over a discrete $(d_P,\,v)$ grid, the expected number of impacts as a function of crater depth and aperture diameter is derived. The impact velocity distribution $\phi(v)$ is modeled as log-normal with median $\bar{v} = \SI{17}{km/s}$ and logarithmic standard deviation $\sigma_{\ln v} = 0.45$~\cite{taylor1995meteoroid}. The result for a ten-year mission duration is shown in Fig.\,\ref{fig:ImpactCraterSizeDistribution}.

For the impact damage to remain self-healing, the crater depth must not exceed the local liquid column depth $h_0$, i.e.\ $d_P \leq h_0$, so that the free surface is not fully punctured and the liquid body remains continuous (besides the loss of a small amount of vaporized liquid that could be replenished). This adds a constraint on the minimum film thickness that is competing with the maximum imposed by setting a limit on the areal density, but leaves a valid design space, since the impact craters are comparably shallow (see Fig.~\ref{fig:ImpactCraterSizeDistribution}). In this regime, the perturbation is a localized surface deformation of lateral extent $2d_P$, which is small relative to the aperture and contributes only marginally to the aperture-averaged RMS surface error. This is accentuated by the sub-threshold regions in Fig.~\ref{fig:ImpactCraterSizeDistribution}, which span across large domains, and impact defects remain below optical limits for both NIR and VIS, far below the one crater per mission duration threshold. Impact events that produce craters large enough to push the RMS surface error beyond optical limits are consequently rare and unlikely to occur over a 10-year mission. Multiple smaller impact events could, in theory, result in significant deformations, depending on the settling timescale.

While the global free-surface dynamics of the spherical LMT film are well described by thin-film theory, the post-impact hemispherical perturbation has a depth-to-radius ratio of order one, violating the small-gradient assumption locally and placing the crater dynamics in the capillary-inertial regime. With the Ohnesorge number~\cite{Ohnesorge1936} $\mathrm{Oh} = \mu/\sqrt{\rho_L\,\gamma\,d_P} \sim \mathcal{O}(10^{-3}) \ll 1$, viscous dissipation is negligible during the initial healing transient. The crater heals on the capillary-inertial timescale~\cite{Rayleigh1879,Eggers_2008} $\tau_{ci} = \sqrt{\rho_L d_P^3/\gamma} \sim \mathcal{O}(0.1) \text{ to } \mathcal{O}(100)$~ms and the resulting capillary oscillations decay viscously~\cite{Lamb1932} on $\tau_{visc} = \rho_L d_P^2/(2\mu) \sim \mathcal{O}(0.1) \text{ to } \mathcal{O}(100)$~s for crater depths in the range 0.1–10 mm shown in Fig.~\ref{fig:ImpactCraterSizeDistribution}, after which the mirror surface returns to its equilibrium shape. Consequently, mirror impacts of significant magnitude are rare and RMS surface error due to the pre-settling impact crater remains below optical limits for non-remote impact events. Even larger craters, which are rather uncommon, heal on timescales of seconds to minutes, well within operational requirements. Self-healing of the liquid mirror surface following micrometeoroid impact is therefore viable and presents a distinct advantage of liquid mirrors.

\subsection{Mirror Control Strategies}
\label{sec:resdis/counteracting}

The preceding sections establish two complementary limits on passively operated, capillary-driven liquid mirror telescopes. In the static regime, solar radiation pressure and self-gravitation induce equilibrium deformations exceeding the $\lambda/20$ tolerance above approximately 0.86~m aperture for visible and 1.58~m for near-infrared observations. In the dynamic regime, sinusoidal filling residuals take years to settle at large apertures, and propulsive maneuvers breach optical limits within seconds to hours, depending on the thrust level. Pure capillary stabilization is therefore insufficient for large apertures, and additional interface control is required both to counteract orbital disturbances and to enable controlled formation of the optical figure. The following subsections quantify the control authority required and assess the feasibility of candidate strategies.

\subsubsection{Counteracting Body Force}
\label{sec:resdis/conservativecontrol}

\begin{figure}[t!]
    \centering
    \includegraphics[width=\linewidth]{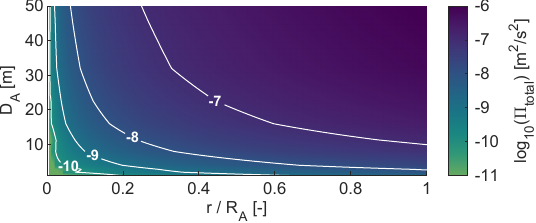}
    \caption{Total perturbation potential $\Pi_{\text{total}}$ (SRP at $\beta=0$°, tidal worst-case at SEL2, point-mass bus, structural and fluid self-gravitation) acting on an unperturbed spherical Galinstan interface as a function of normalized radius $r/R_A$ and aperture diameter $D_A$ for a film thickness $h_0 = 3$\,mm.}
    \label{fig:ControlPotential}
\end{figure}

Two distinct strategies exist to limit interface deformation using conservative body forces: nullifying the perturbation potential through a matched opposing field, or overpowering it by imposing a dominant surface-normal body force that renders the interface insensitive to residual disturbances.

The first strategy introduces a control potential $\Pi_{\text{ctrl,req}} = -\Pi_{\text{total}}$, matching the disturbance field spatially to hold the net potential constant over the spherical profile. Figure~\ref{fig:ControlPotential} represents the sum of the dominant perturbations $\Pi_{\text{total}}$ as a function of the normalized radius $r/R_A$, biased to $\Pi_{\text{total}}(r=0)=0$. The spatial structure of the disturbance field paired with the need for active adjustment to react to spacecraft reorientation rules out uniform body forces, such as axial thrust or spin-induced centrifugal acceleration, as stand-alone correction mechanisms. Dielectrophoretic forces acting on the liquid volume~\cite{dimarco2009} or magnetic body forces acting on a magnetic fluid~\cite{Rosensweig} can in principle produce spatially structured fields tuned to the disturbance potential. However, as seen in Fig.~\ref{fig:ControlPotential}, control potentials of order $\mathcal{O}(10^{-7})$--$\mathcal{O}(10^{-6})$\,m$^2$\,s$^{-2}$ are required at large apertures, while the permissible residual lies below $\mathcal{O}(10^{-8}$--$10^{-9})$\,m$^2$\,s$^{-2}$, imposing a dynamic range requirement that significantly complicates the design of a matched conservative control system.

The second strategy avoids precise potential matching by imposing a strong surface-normal acceleration, leaving the capillary-dominated regime. In the body-force-dominated limit, the equilibrium interface follows the net-potential equipotential, so a steeper surface-normal gradient reduces the displacement produced by a given perturbation potential, effectively stiffening the interface against disturbances. This strategy also presents a direct path to parabolic mirror geometries by shaping the equipotential accordingly.

The Defense Advanced Research Projects Agency (DARPA) Zenith program explored a variety of mirror control approaches for tiltable terrestrial LMTs~\cite{zenith_overview_2024}, with proposed architectures including permanent Halbach arrays~\cite{Rowlands2024_SelfAssemblingILM} and electromagnetic coil arrays combined with Helmholtz coils~\cite{Wu2013} to produce a large-scale, approximately interface-conformal force field. While these systems are designed for Earth-based operation, the underlying actuation concepts are applicable to space, since the forces counteracted are of much greater magnitude than orbital perturbations. The Halbach-based magnetic architecture employed by one Zenith program team is a natural candidate for operating in this body-force-dominated regime, provided a sufficiently uniform surface-normal field can be sustained at scale.

\begin{figure}[t!]
    \centering
    \includegraphics[width=\linewidth]{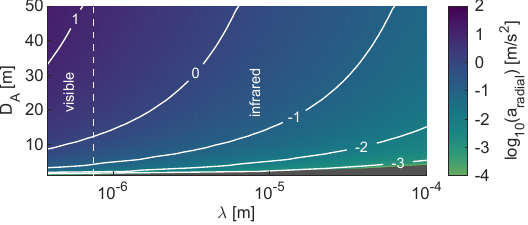}
    \caption{Surface-normal (radial) acceleration required to suppress Galinstan interface deformations below the $\lambda/20$ optical limit, as a function of aperture diameter $D_A$, for representative wavelengths spanning the visible and infrared regimes, a film thickness of $h_0 = 3$\,mm, and $\beta=0$° SRP. The dark gray region denotes configurations already within tolerance without an applied body force. }
    \label{fig:StrongRadialAccel}
\end{figure}

By imposing a surface-normal acceleration and sweeping its magnitude over the aperture range at constant film thickness $h_0=3$\,mm, a relation between the imposed acceleration level and the residual interface deformation under the dominant perturbations (SRP, structural self-gravitation) can be derived. Figure~\ref{fig:StrongRadialAccel} shows the acceleration required to suppress interface deformations below $\lambda/20$ as a function of aperture. The dark gray region denotes configurations that already satisfy the optical requirement without any additional body force. Required levels range from 39~mm\,s$^{-2}$ at 5~m apertures and near-infrared wavelengths up to approximately 16.9\,m\,s$^{-2}$ for a 50\,m aperture mirror at the short end of the visible spectrum. Because SRP depends on spacecraft attitude, the axisymmetric $\beta=0$° case shown here represents a nominal rather than worst-case load; slightly higher accelerations may be required depending on the operational envelope. Designing to the worst-case perturbation, however, yields a fully passive control strategy requiring no active adjustment for attitude variations. Additionally, a hybrid approach consisting of an undersized static radial acceleration supplemented by an active correction that compensates the remaining time-varying perturbation would be conceivable.

\subsubsection{Acoustic Interface Control}
\label{sec:resdis/acoustic}

Acoustic radiation pressure provides a normal-stress route to interface control. Using the interchangeability of the mass-force potential and applied normal stresses, the required stress follows from Fig.~\ref{fig:ControlPotential} directly as $p_\text{req} = \rho_L \Pi_\text{total} + C_n$, with the arbitrary offset $C_n$, leading to normal stresses of order $\mathcal{O}(10^{-4})$--$\mathcal{O}(10^{-3})$~Pa.
Because the mirror operates in vacuum, the acoustic field can only be delivered through the liquid from the substrate. For a liquid--vacuum interface, the radiation stress scales as $p_\text{rad}\sim p_i^{2}/\rho c_L^{2}$~\cite{Borgnis1953}, with $p_i$ the incident acoustic pressure amplitude and $c_L = 2730$~m\,s$^{-1}$ the speed of sound in Galinstan~\cite{Wang2017}, corresponding to acoustic pressure amplitudes of order $\mathcal{O}(10^{3})$--$\mathcal{O}(10^{4})$~Pa.

However, several factors call the viability of this route into question. The most restrictive is cavitation: the drive places the liquid under transient tension of the order of $p_i$ against a near-zero ambient pressure. While no cavitation threshold has been measured for Galinstan, in experiments a thin layer of Ga--In has been shown to cavitate under ultrasonic drive~\cite{Li2021}. The tension a candidate liquid can sustain thus becomes a further strict materials requirement on an already constrained choice of working fluid. Spatially resolving the control field across a 50~m aperture would further require a transducer array orders of magnitude larger than the 256 elements of the largest laboratory demonstration~\cite{Nagakura2025}, itself performed at far smaller scale and greater film thickness. Additionally, operating such an array would inject micro-vibration into the structure. Acoustic actuation therefore cannot presently be classified as a viable control strategy, although its independence from the electromagnetic properties of the working fluid makes it worth investigating.

\subsubsection{Thermocapillary Interface Control}
\label{sec:resdis/thermocapillary}

Thermocapillary actuation imposes temperature gradients to generate Marangoni stresses that redistribute liquid mass along the interface, driving it toward a target shape through a dynamic balance between disturbance-driven and thermally-induced thin-film fluxes~\cite{karbalaei_thermocapillarity_2016}. Using $\Pi_{\text{total}}$ as the source term in Eq.~\ref{eq:T_req}, the required control temperature $T_{\text{ctrl}}(r)$ is obtained by direct quadrature and expressed as the rise relative to the axis, $\Delta T(r) = T_{\text{ctrl}}(r) - T_{\text{ctrl}}(0)$.
\begin{figure}[t!]
    \centering
    \includegraphics[width=\linewidth]{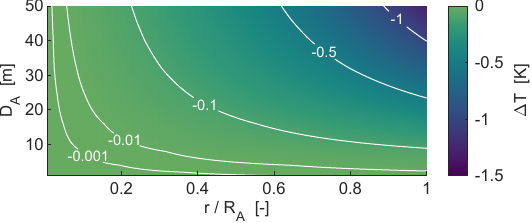}
    \caption{Required temperature difference $\Delta T(r) =T_{\text{ctrl}}(r) - T_{\text{ctrl}}(0)$ for thermocapillary forces to counteract dominant perturbations (SRP at $\beta=0$°, self-gravitation) as a function of normalized radius $r/R_A$ and aperture diameter $D_A$.}
    \label{fig:ThermoCapControl}
\end{figure}
Figure~\ref{fig:ThermoCapControl} maps $\Delta T(r/R_A)$ over aperture diameters from 1 to 50~m at $h_0=3$~mm. The required temperature difference is negative throughout: the negative Marangoni coefficient of Galinstan ($\mathrm{d}\gamma/\mathrm{d}T < 0$) means a warmer center reduces local surface tension, driving outward Marangoni flow that counteracts the net inward perturbation flux. The magnitude grows from approximately $-20$\,mK at $D_A=1$\,m to $-1.5$\,K at $D_A=50$\,m, well within the output range of resistive heater arrays or thermoelectric elements.
\begin{figure}[b!]
    \centering
    \includegraphics[width=\linewidth]{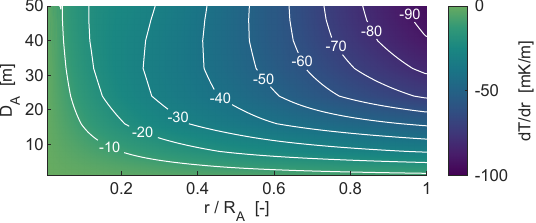}
    \caption{Required temperature gradient $\mathrm{d}T/\mathrm{d}r$ for thermocapillary correction of dominant perturbations (SRP at $\beta=0$°, self-gravitation), as a function of normalized radius $r/R_A$ and aperture diameter $D_A$.}
    \label{fig:ThermoCapControl_Gradient}
\end{figure}
The practical challenge lies not in achieving the required $\Delta T$ magnitude but in maintaining its spatial distribution across the full aperture. Figure~\ref{fig:ThermoCapControl_Gradient} shows the required gradient $\mathrm{d}T/\mathrm{d}r$, which converges to zero at the center and grows toward the rim. Any deviation from the target field redistributes the Marangoni flux and introduces residual interface deformations. 

To assess the practical feasibility of thermocapillary control, a sensitivity analysis is performed for a discrete heater array with node pitch $\Delta r_h$, accounting for (i)~deterministic error from the spatial discretization of $T_\text{ctrl}(r)$ onto a finite heater grid, with piecewise-linear variation assumed between nodes, and (ii)~random per-node heater setpoint errors of standard deviation $\eta$. Linearizing Eq.~\ref{eq:thinfilm_marangoni} about a uniform film of thickness $h_0$, the residual deformation $\delta h$ driven by the temperature mismatch $\Delta T_\text{err}(r) = T_\text{heater}(r) - T_\text{ctrl}(r)$ satisfies
\begin{equation}
    \nabla^2\delta h = \frac{3}{2h_0\gamma_0}\frac{\mathrm{d}\gamma}{\mathrm{d}T}\,\Delta T_\text{err},
    \label{eq:thinfilm_poisson}
\end{equation}
with $\delta h'(0)=0$ and $\delta h(R_A)=0$ due to the boundary conditions. A bisection over $\eta$ identifies the maximum allowable setpoint noise $\eta_{\max}$ such that the 95th-percentile rms surface error $\sigma_h$ across 1000 Monte Carlo realizations equals $\lambda/20$ for each $(\Delta r_h, D_A)$ pair.
\begin{figure}[b!]
    \centering
    \begin{subfigure}{\linewidth}
        \centering
        \includegraphics[width=\linewidth]{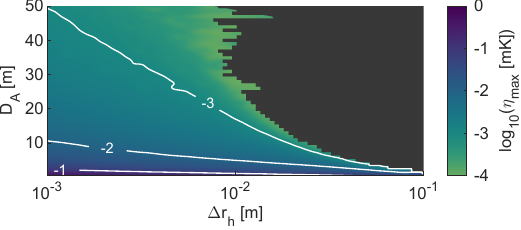}
        \caption{Visible wavelength ($\lambda = 380$~nm).}
    \end{subfigure}
    \vspace{0.5cm}
    \begin{subfigure}{\linewidth}
        \centering
        \includegraphics[width=\linewidth]{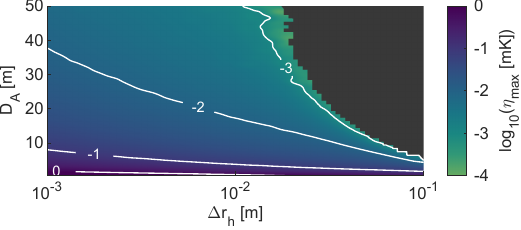}
        \caption{Near-infrared wavelength ($\lambda = 2.5~\mu$m).}
    \end{subfigure}
    \caption{Maximum allowable heater setpoint noise $\eta_{\max}$ as a function of heater pitch $\Delta r_h$ and aperture diameter $D_A$. Dark regions mark heater pitches at which discretisation error alone exceeds the $\lambda/20$ optical tolerance. The speckled appearance within the coloured region reflects Monte Carlo variance in the 95th-percentile $\sigma_h$ estimate across 1000 realisations.}
    \label{fig:NoiseBudget}
\end{figure}

Figure~\ref{fig:NoiseBudget} maps $\eta_{\max}$ for visible ($\lambda=380$~nm) and near-infrared ($\lambda=2.5\,\mu$m) tolerances. Dark regions indicate where deterministic discretization error alone exceeds $\lambda/20$, establishing a hard lower bound on heater pitch independent of noise level. In the NIR band, apertures up to 8.0~m and 3.8~m are accessible at $\eta_{\max} < 0.1$~mK for $\Delta r_h = 1$~mm and 10~mm respectively, extending to 37.7~m and 17.3~m at $\eta_{\max} = 10\,\mu$K. The visible band is more demanding but remains viable at sub-millikelvin levels: at $\eta_{\max} = 0.1$~mK, apertures up to 2.3~m and 1.1~m are accessible for $\Delta r_h = 1$~mm and 10~mm respectively, rising to 10.6~m and 5.0~m at $\eta_{\max} = 10\,\mu$K.

Comparing these requirements against the state of the art in precision thermal control, NIR thermocapillary correction appears practically feasible for multi-meter apertures with current hardware. Closed-loop resistive heater systems have demonstrated better than 2~mK RMS in thermal-vacuum testing of a 1.5~m, 25-zone mirror heater assembly at NASA MSFC~\cite{Brooks2021,Brooks2022}, where the measurement floor was limited by the 22-bit data acquisition system rather than actuator physics~\cite{Brooks2021}, already enabling NIR apertures approaching 2~m. Sub-millikelvin performance, achievable through higher-resolution analogue-to-digital conversion and increased zone heat capacitance, would extend NIR operation beyond 8~m and open visible-wavelength thermocapillary control to apertures of a few meters. Reaching $\eta \approx 10\,\mu$K would demand dedicated precision thermal hardware~\cite{Havey2019} but could enable thermocapillary correction across ten or more meters in aperture for both wavelength bands. 

More broadly, the noise budget established in Fig.~\ref{fig:NoiseBudget} constitutes a general thermal uniformity requirement for any liquid mirror concept, not only those employing thermocapillary actuation as interface control method. Any temperature non-uniformity across the aperture drives Marangoni flows through Eq.~\ref{eq:thinfilm_marangoni} and distorts the optical surface. Therefore, the same $\eta_{\max}$ limits apply as passive thermal design constraints for capillary-dominated LMTs: near-infrared LMTs must maintain aperture-wide temperature uniformity at the mK to sub-mK level, and visible-wavelength operation demands uniformity at the $1-100~\mu$K level. Under these constraints, the contribution of thermal expansion to the error budget is secondary and can be addressed through appropriate material selection. Enforcing strong interface-normal accelerations would dampen these induced thermocapillary fluxes and thus relax the temperature requirements.

\subsubsection{Segmentation} \label{sec:resdis/segmentation}
In analogy to the segmented primary mirror of the James Webb Space Telescope (JWST)~\cite{Gardner2006}, segmentation is an alternative path to achieving large equivalent apertures with purely capillary-driven mirrors. The central motivation follows directly from the static equilibrium results presented in Sec.~\ref{sec:resdis/Static_Response}, where small-aperture mirrors satisfy the $\lambda/20$ optical criterion without the aforementioned control mechanisms. Segmentation also recovers favorable dynamic behavior since individual segments recover from perturbations and complete initial formation orders of magnitude faster than a monolithic 50~m mirror. The aperture limits derived in Sec.~\ref{sec:resdis/Static_Response} can, however, not be taken directly as the maximum segment size for a larger-scale, purely capillary-dominated, segmented architecture, since not all perturbations scale independently of the total telescope size. The SRP acceleration depends on the area-to-mass ratio of the full system, and the bus mass contribution diminishes monotonically as total mirror area grows, causing the acceleration to converge toward the pure-mirror limit. The maximum passively compliant segment aperture, therefore, decreases with increasing target aperture, and the single-mirror thresholds represent only lower bounds on segment count.

Consequently, a segmented architecture transforms the aperture-scaling problem into a formation and co-phasing problem. Segment-to-segment piston errors must be controlled to a fraction of the observing wavelength via active wavefront sensing~\cite{Chanan2000}, and tip-tilt errors must be held within a fraction of the diffraction-limited resolution. Gaps between segments additionally reduce fill factor and introduce point-spread function artifacts absent in a monolithic design~\cite{Troy2003}.

\begin{table*}[t!]
\centering
\caption{Performance metrics for the purely capillary regime and control
strategies for representative 0.8~m, 9~m and 50~m aperture Galinstan LMTs at
visible ($\lambda = 380$\,nm) and near-infrared ($\lambda = 2500$\,nm)
wavelengths, with film thickness $h_0 = 3$~mm in SEL2.}
\label{tab:representative_cases}

\setlength{\tabcolsep}{8pt}
\begin{tabular*}{\linewidth}{@{\extracolsep{\fill}} l c cccccc c}
\toprule
\multirow{2}{*}{\text{Performance Metric}}
  & \text{Aperture [m]}
  & \multicolumn{2}{c}{\text{0.8}}
  & \multicolumn{2}{c}{\text{9}}
  & \multicolumn{2}{c}{\text{50}}
  & \multirow{2}{*}{\text{Units}} \\
\cmidrule(lr){2-2}\cmidrule(lr){3-4}\cmidrule(lr){5-6}\cmidrule(lr){7-8}
  & \text{Wavelength [nm]}
  & \multicolumn{1}{c|}{\text{380}} & \text{2500}
  & \multicolumn{1}{c|}{\text{380}} & \text{2500}
  & \multicolumn{1}{c|}{\text{380}} & \text{2500}
  & \\
\midrule
\multicolumn{9}{l}{\textbf{Purely Capillary LMT}} \\
RMS surface error
  & & \multicolumn{2}{c}{13.3} & \multicolumn{2}{c}{$73.9\times10^{3}$}
  & \multicolumn{2}{c}{$44.7\times10^{6}$} & \U{nm} \\
Post-fill settling time
  & & 16.8~min & 12.7~min & 0.864 & 0.599 & 581 & 339 & \U{mos.} \\
Response time, 1\,N thruster
  & & 1.05 & 9.08 & 5.78 & 47.6 & 152 & 1240 & \U{ms} \\
Response time, 1\,mN thruster
  & & \multicolumn{2}{c}{sub-threshold} & 15.9 & 135 & 408 & 3390 & \U{s} \\ 
\addlinespace[6pt]
\multicolumn{9}{l}{\textbf{Control Strategies}} \\ 
Req.\ surf.-norm.\ acceleration
  & & \multicolumn{2}{c}{not required} & 1.09 & 0.155 & 16.9 & 2.55
  & \U{m\,s$^{-2}$} \\
Max therm.\ noise, $\Delta r_h = 10$\,mm
  & & \multicolumn{2}{c}{153} & 3.87 & 27.0 & N/A & 1.39
  & \U{$\mu$K} \\
Max therm.\ noise, $\Delta r_h = 1$\,mm
  & & \multicolumn{2}{c}{480} & 13.0 & 85.5 & 0.982 & 6.14
  & \U{$\mu$K} \\
\bottomrule
\end{tabular*}
\end{table*}

At the single-mirror thresholds, a 50~m aperture already requires on the order of $1{,}500$ hexagonal 1~m segments, with actual counts exceeding this for large arrays. Holding them in alignment demands either a rigid truss or formation flying with per-segment propulsion and GNC. A complication specific to liquid mirrors is that each container rim constitutes an interior contact line of the combined aperture, so rim-height non-uniformities propagate deformations into the optically active region rather than to the periphery, tightening manufacturing tolerances on individual segment containers. Taken together, segmentation is conceptually attractive but shifts the design burden toward precision formation maintenance, active co-phasing of liquid surfaces, and high segment count.

\subsection{Representative Implementations}
\label{sec:resdis/RepresentativeImplementations}

Two aperture cases bracket the practically relevant design space. At the time of writing, a 9~m aperture represents the largest monolithic mirror that can be delivered within current/near-future launch vehicle fairings~\cite{SpaceX2020Starship} and serves as a direct, rigid-mirror successor to the 6.5~m segmented primary of JWST. A 50~m aperture corresponds to the long-term target defined by the FLUTE program~\cite{Balaban2025FLUTE_Poster}, which is motivated by the astrophysics community's need for substantially larger light-collecting areas across a broad range of science cases~\cite{LUVOIR_2019, Astro2020}.

Performance metrics for these two implementations, as well as a mirror operating within the capillary limit at $D_A = 0.8$~m, are summarized in Table~\ref{tab:representative_cases}. A film thickness of $h_0 = 3$~mm is assessed throughout, consistent with the areal density of JWST. Both $D_A = 9$~m and $D_A = 50$~m lie far beyond the capillary limit under the $\lambda/20$ criterion ($\lambda = 380$~nm: ${D_A}_\text{max} = 0.86$~m; $\lambda = 2.5\,\mu$m: ${D_A}_\text{max} = 1.58$~m), with RMS surface errors of 73.9~$\mu$m and 44.7~mm, respectively. Mirror control systems are therefore mandatory for both implementations. For the 50~m aperture, surface control is additionally required to reduce post-fill settling timescales, which otherwise exceed decades. Thermocapillary control via surface heaters is feasible for the 9~m NIR case within current thermal actuator capabilities, while improvements in heater precision are required for the 9~m visible system. For the 50~m mirror, the body-force-dominated control strategy is required, with surface-normal accelerations ranging from 16.9~m\,s$^{-2}$ for visible observations to 2.55~m\,s$^{-2}$ for near-infrared.

\section{Conclusions} \label{sec:conclusions}

Self-gravitation and solar radiation pressure are found to be the dominant perturbations in SEL2 orbits, leading to significant deformations of the optical interface for large-scale capillary-dominated liquid mirror telescopes. Liquid deformations exceed optical limits for visible light (380~nm) at apertures of 0.86~m and 1.58~m for NIR ($\lambda = 2.5$~um). Capillary LMT dynamics are favorable for smaller deformations, exhibiting quick self-healing abilities after particle impacts ($\mathcal{O}(0.1) \text{ to } \mathcal{O}(100)$~s). Although this is a clear advantage of the liquid mirror architecture, dynamics at the entire aperture scale quickly become prohibitive: post-filling settling requires years to decades to relax to optical tolerances for apertures approaching 50~m, and propulsive maneuvers breach optical limits within seconds under chemical-thruster forcing and within hours under low-thrust electric propulsion. Increasing film thickness accelerates settling but raises areal density and amplifies the transient response to thrust. Galinstan outperforms ionic liquid candidates by nearly three orders of magnitude in settling time, driven by its lower viscous--capillary ratio $\mu/\gamma$.

These findings establish that passive, capillary-dominated fluid positioning is insufficient for large-aperture LMTs. Additional interface control is required to counteract orbital disturbances and to aid in the response to large-scale transient events. Among the candidate strategies considered in this paper, spatially resolved conservative body-force control is theoretically capable of nullifying perturbations but demands high dynamic range and tight spatial fidelity, significantly complicating implementation. Thermocapillary actuation via resistive heater arrays allows apertures up to 2.3~m at $100~\mu$K and 10.6~m at $10~\mu$K resolution for visible observations, and up to 8.0~m and 37.7~m, respectively, in the near-infrared, with sub-millikelvin performance achievable using current hardware and $10~\mu$K demanding dedicated precision thermal systems. More broadly, the derived temperature uniformity requirements constitute a passive thermal design constraint applicable to all space LMT concepts: any aperture-wide temperature non-uniformity drives Marangoni flows that distort the optical surface regardless of the chosen control strategy, imposing mK-to-sub-mK uniformity requirements for near-infrared operation and $1-100~\mu$K uniformity for visible wavelengths. Acoustic radiation pressure offers an alternative route, but the cavitation resistance of the working liquid, the introduction of vibrations into the entire system, and the scale of the required transducer array undermine its viability. Leaving the capillary-dominated regime by imposing a strong surface-normal acceleration, for instance through magnetic body forces on a ferrofluid mirror, emerges as the most promising concept for very large space LMTs. This settling force can passively suppress perturbation sensitivity across the full aperture without requiring active adjustment for varying spacecraft orientation, opens a direct path to parabolic mirror geometries, and partially dampens thermocapillary fluxes, thereby relaxing the thermal uniformity requirements. Segmentation offers an alternative purely capillary path by keeping individual mirror diameters within the passively compliant regime, but transforms the aperture-scaling problem into one of precision formation maintenance, active co-phasing of liquid surfaces, and high segment count.

\section*{CRediT authorship contribution statement}
\textbf{Janoah Dietrich:} Writing – review \& editing, Writing – original draft, Visualization, Methodology, Validation, Software, Investigation, Data curation.
\textbf{{\'A}lvaro {Romero-Calvo}:} Writing – review \& editing, Supervision, Project administration, Resources, Methodology, Conceptualization, Investigation.

\section*{Declaration of competing interest}
The authors declare that they have no known competing financial interests or personal relationships that could have appeared to influence the work reported in this paper.

\section*{Acknowledgments}
The authors thank Dr. Michael “Orbit” Nayak with the Defense Advanced Research Projects Agency (DARPA) for the initial discussions leading to this work. Comments and feedback from Mr. Filippo di Benedetto, Prof. Ralf Srama, Prof. Gabriel Cano-Gómez, Prof. Miguel Herrada, Dr. Neil Rowlands, Dr. Amanda Childers, and Mr. David Strafford are gratefully acknowledged. The simulations were partially supported through research cyberinfrastructure resources and services provided by the Partnership for an Advanced Computing Environment (PACE) at the Georgia Institute of Technology, Atlanta, Georgia, USA.


\bibliographystyle{elsarticle-num.bst}

\bibliography{references}

@article{nasaDeploymentExplorer,
	title        = {The James Webb Space Telescope Mission: optical telescope element design, development, and performance},
	author       = {McElwain, Michael W. and Feinberg, Lee D. and Perrin, Marshall D. and Clampin, Mark and Mountain, C. Matt and Lallo, Matthew D. and Lajoie, Charles-Philippe and Kimble, Randy A. and Bowers, Charles W. and Stark, Christopher C.},
	year         = 2023,
	journal      = {Publications of the Astronomical Society of the Pacific},
	volume       = 135,
	number       = 1047,
	pages        = {058001},
	doi          = {10.1088/1538-3873/acada0 }
}

@article{Park2023,
  author    = {Park, S. and Liu, L. and Demirkir, C. and van der Heijden, O. and Lohse, D. and Krug, D. and Koper, M. T. M.},
  title     = {Solutal {Marangoni} effect determines bubble dynamics during electrocatalytic hydrogen evolution},
  journal   = {Nature Chemistry},
  year      = {2023},
  volume    = {15},
  pages     = {1532--1540},
  doi       = {10.1038/s41557-023-01294-y}
}

@article{Flute2,
	title        = {Shaping a gallium alloy and an ionic liquid into spherical mirrors for future liquid-based telescopes-experimental setup and demonstration in parabolic flights},
	author       = {Luria, Omer and Gommed, Khaled and Elgarisi, Mor and Gabay, Israel and Ericson, Jonathan and Frumkin, Valeri and Razin, Alexey and Widerker, Daniel and Belikov, Ruslan and Bookbinder, Jay},
	year         = 2024,
	journal      = {Journal of Astronomical Telescopes, Instruments, and Systems},
	volume       = 10,
	number       = 4,
	pages        = {044010},
	doi          = {10.1117/1.JATIS.10.4.044010 }
}

@article{postman2012advanced,
	author       = {Postman, Marc and Brown, Tom and Sembach, Kenneth and Giavalisco, Mauro and Traub, Wesley and Stapelfeldt, Karl and Calzetti, Daniela and Oegerle, William and Rich, R. Michael and Stahl, H. Philip},
	title        = {Advanced technology large-aperture space telescope: science drivers and technology developments},
	journal      = {Optical Engineering},
	year         = 2012,
	volume       = 51,
	number       = 1,
	pages        = {011007},
	doi          = {10.1117/1.OE.51.1.011007 }
}

@article{photonics12030199,
	title        = {A Review of Exoplanet Detection Telescopes: Performance Design and Technology Optimization},
	author       = {Sun, Rui and An, Qichang and Wu, Xiaoxia},
	year         = 2025,
	journal      = {Photonics},
	volume       = 12,
	number       = 3,
	pages        = 199,
	doi          = {10.3390/photonics12030199 }
}

@book{Schiesser1991,
  author    = {Schiesser, William E.},
  title     = {The Numerical Method of Lines: Integration of Partial Differential Equations},
  publisher = {San Diego},
  year      = {1991},
  isbn      = {978-0126241303}
}

@book{Marov2015,
	author       = {Marov, Mikhail Ya.},
	year         = 2015,
	title    = {The Fundamentals of Modern Astrophysics: A Survey of the Cosmos from the Home Planet to Space Frontiers},
	publisher    = {Springer},
    address      = {New York},
	pages        = {279--294},
	isbn         = {978-1-4614-8729-6},
    doi          = {10.1007/978-1-4614-8730-2}
}

@article{Stahl2010Survey,
  author    = {Stahl, H. Philip},
  title     = {Survey of cost models for space telescopes},
  journal   = {Optical Engineering},
  volume    = {49},
  number    = {5},
  pages     = {053005},
  year      = {2010},
  publisher = {SPIE},
  doi       = {10.1117/1.3430603},
}

@article{Riess2022,
  author    = {Riess, Adam G. and Yuan, Wenlong and Macri, Lucas M. and Casertano, Stefano and Scolnic, Dan},
  title     = {A Comprehensive Measurement of the Local Value of the Hubble Constant with 1 km s-1 Mpc-1 Uncertainty from the Hubble Space Telescope and the SHOES Team},
  journal   = {The Astrophysical Journal Letters},
  volume    = {934},
  number    = {1},
  pages     = {L7},
  year      = {2022},
  doi       = {10.3847/2041-8213/ac5c5b}
}

@article{Planck2020,
  author    = {Planck Collaboration and Aghanim, N. and Akrami, Y.},
  title     = {Planck 2018 results. VI. Cosmological parameters},
  journal   = {Astronomy \& Astrophysics},
  volume    = {641},
  pages     = {A6},
  year      = {2020},
  doi       = {10.1051/0004-6361/201833910}
}

@article{Shapley2011,
  author    = {Shapley, Alice E.},
  title     = {Physical Properties of Galaxies from z=2-4},
  journal   = {Annual Review of Astronomy and Astrophysics},
  volume    = {49},
  pages     = {525--580},
  year      = {2011},
  doi       = {10.1146/annurev-astro-081710-102542}
}

@article{Koekemoer2011,
  author    = {Koekemoer, Anton M. and Faber, S. M. and Ferguson, H. C.},
  title     = {CANDELS: THE COSMIC ASSEMBLY NEAR-INFRARED DEEP EXTRAGALACTIC LEGACY SURVEY-THE
HUBBLE SPACE TELESCOPE OBSERVATIONS, IMAGING DATA PRODUCTS, AND MOSAICS},
  journal   = {The Astrophysical Journal Supplement Series},
  volume    = {197},
  number    = {2},
  pages     = {36},
  year      = {2011},
  doi       = {10.1088/0067-0049/197/2/36}
}

@article{Meszaros2006,
  author    = {M{\'e}sz{\'a}ros, P{\'e}ter},
  title     = {Gamma-ray bursts},
  journal   = {Reports on Progress in Physics},
  volume    = {69},
  number    = {8},
  pages     = {2259--2321},
  year      = {2006},
  doi       = {10.1088/0034-4885/69/8/R01}
}

@article{Wilms2001,
    author = {Lumsden, S.L. and Alexander, D.M.},
    title = {The infrared luminosity of the torus and the visibility of scattered broad line emission in Seyfert 2 galaxies},
    journal = {Monthly Notices of the Royal Astronomical Society},
    volume = {328},
    number = {3},
    pages = {L32-L36},
    year = {2001},
    month = {12},
    doi = {10.1046/j.1365-8711.2001.05074.x},
}

@article{Scoville2007,
  author    = {Scoville, N. and Abraham, R. G. and Aussel, H.},
  title     = {COSMOS: Hubble Space Telescope Observations},
  journal   = {The Astrophysical Journal Supplement Series},
  volume    = {172},
  number    = {1},
  pages     = {38--45},
  year      = {2007},
  doi       = {10.1086/516580}
}

@inproceedings{Collicott2007,
  author    = {Steven H. Collicott},
  title     = {Impact of Non-uniform Acceleration Fields on Future Spacecraft with Large Tanks},
  booktitle = {43rd AIAA/ASME/SAE/ASEE Joint Propulsion Conference \& Exhibit},
  year      = {2007},
  publisher = {AIAA},
  doi       = {10.2514/6.2007-5554},
  pages     = {1--9}
}

@article{Caupin2012,
  author       = {Fr{\'e}d{\'e}ric Caupin and Arnaud Arvengas and Kristina Davitt and Mouna El Mekki Azouzi and Kirill I. Shmulovich and Claire Ramboz and David A. Sessoms and Abraham D. Stroock},
  title        = {Exploring water and other liquids at negative pressure},
  journal      = {Journal of Physics: Condensed Matter},
  volume       = {24},
  number       = {28},
  pages        = {284110},
  year         = {2012},
  doi          = {10.1088/0953-8984/24/28/284110},
}

@article{RomeroCalvo2021_AxisymmetricFerrofluid,
  author       = {Romero-Calvo, {\'A}lvaro and Herrada, Miguel Angel and Hermans, Tim H. J. 
                  and Parrilla Ben{\'\i}tez, Lidia and Cano-G{\'o}mez, Gabriel and Castro-Hernández, Elena},
  title        = {Axisymmetric Ferrofluid Oscillations in a Cylindrical Tank in Microgravity},
  journal      = {Microgravity Science and Technology},
  volume       = {33},
  pages        = {50},
  year         = {2021},
  doi          = {10.1007/s12217-021-09894-4}
}

@article{RomeroCalvo2021_MagneticPositivePositioning,
title = {Magnetic Positive Positioning: Toward the application in space propulsion},
journal = {Acta Astronautica},
volume = {187},
pages = {348-361},
year = {2021},
issn = {0094-5765},
doi = {10.1016/j.actaastro.2021.06.045},
author = {{\'A}lvaro Romero-Calvo and F. Maggi and H. Schaub},
}

@book{Wu2013_MFDM_Book,
  author    = {Wu, Zhizheng and Iqbal, Azhar and Ben Amara, Foued},
  title     = {Modeling and Control of Magnetic Fluid Deformable Mirrors for Adaptive Optics Systems},
  year      = {2013},
  publisher = {Springer},
  address   = {Berlin, Heidelberg},
  series    = {Springer Series in Optical Sciences},
  isbn      = {978-3-642-32228-0},
  doi       = {10.1007/978-3-642-32229-7},
  pages     = {81--83}
}

@article{Zheng2018_HealingCapillaryFilms,
  title        = {Healing capillary films},
  author       = {Zheng, Zhong and Fontelos, Marco A. and Shin, Sangwoo and Dallaston, Michael C. and Tseluiko, Dmitri and Kalliadasis, Serafim and Stone, Howard A.},
  journal      = {Journal of Fluid Mechanics},
  volume       = {838},
  pages        = {404--434},
  year         = {2018},
  doi          = {10.1017/jfm.2017.777}
}

@article{BahcallSpitzer1982,
	title        = {The Space Telescope},
	author       = {Bahcall, John N. and Spitzer, Lyman},
	year         = 1982,
	journal      = {Scientific American},
	volume       = 247,
	number       = 1,
	pages        = {40--51},
    url          = {https://www.jstor.org/stable/24966634}
}

@inproceedings{2011Trova_SelfGrav,
  author       = {Trova, A. and Hur{\'e}, J.-M. and Hersant, F.},
  title        = {The gravitational potential of axially symmetric bodies from a regularized green kernel},           
  booktitle    = {{SF2A-2011}: Proceedings of the Annual Meeting of the French Society of Astronomy and Astrophysics},
  year         = {2011},
  pages        = {685--688},    
  address      = {Paris, France},
  publisher    = {Soci{\'e}t{\'e} Francaise d'Astronomie et d'Astrophysique},
  URL          = {https://sf2a.eu/proceedings/2011/2011sf2a.conf..0685T.pdf}
}

@book{Schaub2018,
  author    = {Schaub, Hanspeter and Junkins, John L.},
  title     = {Analytical Mechanics of Space Systems},
  year      = {2018},
  publisher = {American Institute of Aeronautics and Astronautics},
  address   = {Reston, Virginia},
  series    = {AIAA Education Series},
  edition   = {4th},
  isbn      = {978-1-62410-521-0},
  doi       = {10.2514/4.105210},
  pages     = {767--777},
}

@book{Vallado2013,
  author    = {Vallado, David A.},
  title     = {Fundamentals of Astrodynamics and Applications},
  year      = {2013},
  publisher = {Microcosm Press},
  address   = {Hawthorne, California},
  series    = {Space Technology Library},
  edition   = {4th},
  isbn      = {978-1-881883-18-0},
  pages     = {388--393},
}

@book{McInnes1999,
  author    = {McInnes, Colin R.},
  title     = {Solar Sailing: Technology, Dynamics and Mission Applications},
  year      = {1999},
  publisher = {Springer},
  address   = {London, UK},
  series    = {Springer Praxis Books},
  isbn      = {978-1-85233-102-1},
}

@book{Gilmore2002,
  author    = {Gilmore, David G.},
  title     = {Spacecraft Thermal Control Handbook, Volume I: Fundamental Technologies},
  year      = {2002},
  publisher = {The Aerospace Press},
  address   = {El Segundo, CA},
  edition   = {2nd},
  isbn      = {978-1-884989-11-7},
}

@article{flywheel_microvibration_2021,
  author  = {Yu, Yang and Gong, Xiaoxue and Zhang, Lei and
             Jia, Hongguang and Xuan, Ming},
  title   = {Full-Closed-Loop Time-Domain Integrated Modeling Method
             of Optical Satellite Flywheel Micro-Vibration},
  journal = {Applied Sciences},
  volume  = {11},
  number  = {3},
  pages   = {1328},
  year    = {2021},
  doi     = {10.3390/app11031328}
}

@inproceedings{Hyde2004,
  author    = {T. T. Hyde and K. Q. Ha and J. D. Johnston and J. M. Howard and G. E. Mosier},
  title     = {Integrated modeling activities for the James Webb Space Telescope: optical jitter analysis},
  booktitle = {Proceedings of SPIE: The International Society for Optical Engineering},
  volume    = {5487},
  pages     = {588--599},
  year      = {2004},
  doi       = {10.1117/12.551806}
}

@inproceedings{Liu2013AIAA_2008_7232,
  author    = {Kuo-Chia Liu and P. Maghami and C. Blaurock},
  title     = {Reaction Wheel Disturbance Modeling, Jitter Analysis, and Validation Tests for Solar Dynamics Observatory},
  booktitle = {Proceedings of the AIAA Guidance, Navigation and Control Conference and Exhibit},
  year      = {2008},
  series    = {AIAA Paper},
  number    = {2008-7232},
  address   = {Honolulu, HI, USA},
  month     = {8},
  doi       = {10.2514/6.2008-7232},
  pages     = {1--18}
}

@techreport{Nelson1994GJitter,
  author      = {Nelson, Emily S.},
  title       = {An Examination of Anticipated g-Jitter on {Space Station}
                 and Its Effects on Materials Processes},
  institution = {NASA Lewis Research Center},
  type        = {Technical Memorandum},
  number      = {NASA-TM-103775},
  address     = {Cleveland, OH, USA},
  year        = {1994},
  month       = sep,
  url         = {https://ntrs.nasa.gov/citations/19950006290}
}

@article{RomeroCalvo2020MagneticLiquids,
  author       = {Romero-Calvo, {\'A}lvaro and Cano-G{\'o}mez, Gabriel and Castro-Hern{\'a}ndez, Elena and Maggi, Filippo},
  title        = {Free and Forced Oscillations of Magnetic Liquids},
  journal      = {Journal of Applied Mechanics},
  volume       = {87},
  number       = {2},
  pages        = {021010},
  year         = {2019},
  doi          = {10.1115/1.4045620},
}

@book{wie2008space,
  author       = {Bong Wie},
  title        = {Space Vehicle Dynamics and Control},
  edition      = {2nd},
  series       = {AIAA Education Series},
  publisher    = {American Institute of Aeronautics and Astronautics},
  address      = {Reston, VA, USA},
  year         = {2008},
  isbn         = {978-1-56347-953-3},
  doi          = {10.2514/4.860119},
}

@inproceedings{LUMIO,
  author       = {A. Romero-Calvo and J. Biggs and F. Topputo},
  title        = {Attitude Control for the LUMIO CubeSat in Deep Space},
  booktitle    = {Proceedings of the 70th International Astronautical Congress (IAC)},
  year         = {2019},
  pages        = {1--13},
  organization = {International Astronautical Federation},
  url          = {https://www.researchgate.net/publication/336778960_Attitude_Control_for_the_LUMIO_CubeSat_in_Deep_Space}
}

@article{RomeroCalvo2022_FerrohydrodynamicInterfaceTracking,
  author  = {Romero-Calvo, {\'A}lvaro and Herrada, Miguel A. and Cano-G{\'o}mez, Gabriel and Schaub, Hanspeter},
  title   = {Fully coupled interface-tracking model for axisymmetric ferrohydrodynamic flows},
  journal = {Applied Mathematical Modelling},
  volume  = {111},
  pages   = {836--861},
  year    = {2022},
  doi     = {10.1016/j.apm.2022.06.046}
}

@article{Gomez2006EMIMES,
	title        = {Physical Properties of Pure 1-Ethyl-3-methylimidazolium Ethylsulfate and Its Binary Mixtures with Ethanol and Water at Several Temperatures},
	author       = {G{\'o}mez, Elena and GonzAlez, Bego{\~n}a and Calvar, Noelia and Tojo, Emilia and Dom{\'\i}nguez, Angeles},
	year         = 2006,
	journal      = {Journal of Chemical \& Engineering Data},
	volume       = 51,
	number       = 6,
	pages        = {2096--2102},
	doi          = {10.1021/je060228n}
}

@article{HANDSCHUHWANG2022101642,
	title        = {The subtle difference between Galinstan (R) and eutectic GaInSn},
	author       = {Handschuh-Wang, Stephan and Gan, Tiansheng and Rauf, Muhammad and Yang, Weifa and Stadler, Florian J. and Zhou, Xuechang},
	year         = 2022,
	journal      = {Materialia},
	volume       = 26,
	pages        = 101642,
	doi          = {10.1016/j.mtla.2022.101642}
}

@article{Pappa2003,
	title        = {Structural Dynamics Experimental Activities in Ultralightweight and Inflatable Space Structures},
	author       = {Pappa, Richard S. and Lassiter, John O. and Ross, Brian P.},
	year         = 2003,
	journal      = {Journal of Spacecraft and Rockets},
	volume       = 40,
	number       = 1,
	pages        = {5--16},
	doi          = {10.2514/2.3934}
}

@book{Nasa_Strategic_Plan,
	title        = {Foresight, Strategy and Futures Studies for Defense and Security},
	author       = {Rodgers, Erica},
	year         = 2021,
	publisher    = {Springer},
    address      = {Cham, Switzerland},
	pages        = {1--16},
    URL          = {https://ntrs.nasa.gov/api/citations/20210013666/downloads/Foresight%20Book%20Chapter_Final%2020211105.docx.pdf}
}

@book{Newton1687,
	title        = {Philosophiae Naturalis Principia Mathematica},
	author       = {Newton, Isaac},
	year         = 1687,
	publisher    = {Joseph Streater for the Royal Society},
	address      = {London},
    pages        = {1-320}
}

@book{Myshkis1987,
	title        = {Low-Gravity Fluid Mechanics: Mathematical Theory of Capillary Phenomena},
	author       = {Myshkis, A. D. and Babskii, Vitalii G. and Kopachevski{\u\i}, N. D. and Slobozhanin, Lev A. and Tiuptsov, A. D.},
	year         = 1987,
	publisher    = {Springer-Verlag},
	address      = {Berlin},
	isbn         = {0387161899},
}

@book{Langbein2002,
	title        = {Capillary Surfaces},
	author       = {Langbein, Dieter},
	year         = 2002,
	publisher    = {Springer},
	address      = {Berlin},
	isbn         = {978-3-540-45267-6},
    doi          = {10.1007/3-540-45267-2}
}

@book{dreyer_free_2007,
	title        = {Free Surface Flows under Compensated Gravity Conditions},
	author       = {Dreyer, Michael},
	year         = 2007,
	publisher    = {Springer-Verlag},
	address      = {Berlin, Heidelberg},
	series       = {Springer Tracts in Modern Physics},
	volume       = 221,
	doi          = {10.1007/3-540-44628-1},
	isbn         = {978-3-540-44626-2}
}

@book{Antar,
	title        = {Fundamentals of Low Gravity Fluid Dynamics and Heat Transfer},
	author       = {Antar, Basil N. and Nuotio-Antar, Vappu S.},
	year         = 1993,
	publisher    = {CRC Press},
	address      = {Boca Raton, Florida},
	isbn         = {9780849389139},
    doi          = {10.1201/9781351072182}
}

@book{Cheng2009,
	title        = {The Principles of Astronomical Telescope Design},
	author       = {Cheng, Jingquan},
	year         = 2009,
	publisher    = {Springer Science+Business Media},
	address      = {New York},
	volume       = 360,
	isbn         = {978-0-387-88790-6},
    chapter      = {5},
    pages        = {309--316},
    doi          = {10.1007/b105475}
}

@book{English2017,
	title        = {Space Telescopes},
	author       = {English, Nathan},
	year         = 2017,
	publisher    = {Springer International Publishing},
    address      = {Cham, Switzerland},
	isbn         = {978-3-319-27812-4},
    chapter      = {1},
    pages        = {1-17},
    doi          = {10.1007/978-3-319-27814-8} 
}

@article{Wood1909,
	title        = {The Mercury Paraboloid as a Reflecting Telescope},
	author       = {Wood, Robert W.},
	year         = 1909,
	journal      = {The Astrophysical Journal},
	volume       = 29,
	pages        = {164--176},
	doi          = {10.1086/141639}
}

@article{Borra1982,
	title        = {The Liquid-Mirror Telescope as a Viable Astronomical Tool},
	author       = {Borra, Ermanno F.},
	year         = 1982,
	journal      = {Journal of the Royal Astronomical Society of Canada},
	volume       = 76,
	number       = 1,
	pages        = {245--256},
    URL          = {https://adsabs.harvard.edu/full/1982JRASC..76..245B}
}

@article{Borra1985,
	title        = {Optical-Shop Testing of Liquid Mirrors},
	author       = {Borra, Ermanno F. and Beauchemin, Mario and Arsenault, Robin and Lalande, Robert},
	year         = 1985,
	journal      = {Publications of the Astronomical Society of the Pacific},
	volume       = 97,
	number       = 591,
	pages        = {454--464},
    doi          = {10.1086/131561}
}

@article{Borra1991,
	title        = {The Case for a Liquid Mirror in a Lunar-Based Telescope},
	author       = {Borra, Ermanno F.},
	year         = 1991,
	journal      = {The Astrophysical Journal},
	volume       = 373,
	pages        = {317--321},
	doi          = {10.1086/170053}
}

@article{Borra1992,
	title        = {Liquid Mirrors: Optical Shop Tests and Contributions to the Technology},
	author       = {Borra, E. F. and Content, R. and Girard, L. and Szapiel, S. and Tremblay, L. M. and Boily, E.},
	year         = 1992,
	journal      = {The Astrophysical Journal},
	volume       = 393,
	pages        = {829--847},
	doi          = {10.1086/171550}
}

@article{Duncan1986,
	title        = {A Note on the History of the Liquid-Mirror Telescope},
	author       = {Olsson-Steel, Duncan},
	year         = 1986,
	journal      = {Journal of the Royal Astronomical Society of Canada},
	volume       = 80,
	number       = 3,
	pages        = {128--133},
    url          = {https://adsabs.harvard.edu/full/1986JRASC..80..128O}
}

@article{Cabanac1998,
	title        = {A Search for Peculiar Objects with the NASA Orbital Debris Observatory 3 Meter Liquid Mirror Telescope},
	author       = {Cabanac, Remi A. and Borra, Ermanno F. and Beauchemin, Mario},
	year         = 1998,
	journal      = {The Astrophysical Journal},
	volume       = 509,
	number       = 1,
	pages        = {309--323},
	doi          = {10.1086/306488}
}

@article{Hickson2007,
	title        = {The Large Zenith Telescope: A 6 m Liquid-Mirror Telescope},
	author       = {Hickson, Paul and Pfrommer, Thomas and Crotts, Arlin and Johnson, Ben and Lanzetta, Kenneth M. and Gromoll, Stefan and Truax, Bruce},
	year         = 2007,
	journal      = {Publications of the Astronomical Society of the Pacific},
	volume       = 119,
	number       = 854,
	pages        = {444--455},
	doi          = {10.1086/517621}
}

@article{Hickson2007a,
	title        = {Image Quality of Liquid-Mirror Telescopes},
	author       = {Hickson, Paul and Racine, Ren{\'e}},
	year         = 2007,
	journal      = {Publications of the Astronomical Society of the Pacific},
	volume       = 119,
	number       = 854,
	pages        = {456--465},
	doi          = {10.1086/517619}
}

@article{Angel2008,
	title        = {A Cryogenic Liquid-Mirror Telescope on the Moon to Study the Early Universe},
	author       = {Angel, J. Roger P. and Worden, Simon P. and Borra, Ermanno F. and Eisenstein, Daniel J. and Foing, Bernard and Hickson, Paul and Josset, Jean-Luc and Ma, Ki Bui and Seddiki, Omar and Sivanandam, Suresh and Thibault, Simon and van Susante, Paul},
	year         = 2008,
	journal      = {The Astrophysical Journal},
	volume       = 680,
	pages        = {1582--1594},
	doi          = {10.1086/588034}
}

@article{Detsis2013,
	title        = {Location Selection and Layout for LB10, a Lunar Base at the Lunar North Pole with a Liquid Mirror Observatory},
	author       = {Detsis, Emmanouil and Doule, Ondrej and Ebrahimi, Aliakbar},
	year         = 2013,
	journal      = {Acta Astronautica},
	volume       = 85,
	pages        = {61--72},
	doi          = {10.1016/j.actaastro.2012.12.004}
}

@article{Pradhan2019,
	title        = {Serendipitous Detection and Size Estimation of Space Debris Using a Survey Zenith-Pointing Telescope},
	author       = {Pradhan, Bikram and Hickson, Paul and Surdej, Jean},
	year         = 2019,
	journal      = {Acta Astronautica},
	volume       = 164,
	pages        = {77--83},
	doi          = {10.1016/j.actaastro.2019.07.008}
}

@article{Hebert2004,
	title        = {Automated Detection of Orbital Debris in Digital Video Data from a Telescope},
	author       = {Hebert, Thomas J. and Africano, John and Stansbery, Gene},
	year         = 2004,
	journal      = {Acta Astronautica},
	volume       = 54,
	number       = 6,
	pages        = {455--461},
	doi          = {10.1016/S0094-5765(03)00171-1}
}

@article{Potter1996,
	title        = {Liquid Metal Mirror for Optical Measurements of Orbital Debris},
	author       = {Potter, Andrew and Mulrooney, Mark},
	year         = 1996,
	journal      = {Acta Astronautica},
	volume       = 38,
	number       = 4,
	pages        = {423--425},
	doi          = {10.1016/0094-5765(96)00058-6}
}

@article{Schauer2020,
	title        = {The Ultimately Large Telescope: What Kind of Facility Do We Need to Detect Population III Stars?},
	author       = {Schauer, Anna T. P. and Drory, Niv and Bromm, Volker},
	year         = 2020,
	journal      = {Astrophys. J.},
	volume       = 904,
	number       = 2,
	pages        = 145,
	doi          = {10.3847/1538-4357/abbc0b}
}

@article{Surdej2018,
	title        = {The 4-m International Liquid Mirror Telescope},
	author       = {Surdej, Jean and Hickson, Paul and Borra, Ermanno and Swings, Jean-Pierre and Habraken, Serge and Akhunov, Talat and Bartczak, Przemyslaw and Chand, Hum and De Becker, Micha{\"e}l and Delchambre, Ludovic and Finet, Fran{\c{c}}ois and Kumar, Brajesh and Pandey, Anil and Pospieszalska, Anna and Pradhan, Bikram and Sagar, Ram and Wertz, Olivier and De Cat, Peter and Denis, Stefan and de Ville, Jonathan and Jaiswar, Mukesh Kumar and Lampens, Patricia and Nanjappa, Nandish and Tortolani, Jean-Marc},
	year         = 2018,
	journal      = {Bulletin de la Soci{\'e}t{\'e} Royale des Sciences de Li{\`e}ge},
	volume       = 87,
	pages        = {68--79},
	doi          = {10.25518/0037-9565.7498}
}

@inproceedings{zenith_overview_2024,
author = {Michael Nayak and Denis Brousseau and Amanda Childers and Tomu Hisakado and Kristyn Kadala and Rebecca Kamire and Yifan Li and Dhanushkodi Mariappan and Greg Radighieri and {\'A}lvaro Romero-Calvo and Neil Rowlands and Paul Schroeder and Grey Tarkenton and Simon Thibault and Devin Vollmer and Santanu Basu and Kaushik Iyer},
title = {{Zenith: DARPA's liquid mirror telescope program}},
volume = {13100},
booktitle = {Advances in Optical and Mechanical Technologies for Telescopes and Instrumentation VI},
organization = {International Society for Optics and Photonics},
publisher = {SPIE},
pages = {1310037},
year = {2024},
doi = {10.1117/12.3020080},
}

@inproceedings{Rowlands2024_SelfAssemblingILM,
	title        = {Development of a Self-assembling Ferrofluidic Ionic Liquid Mirror},
	author       = {Rowlands, Neil and Romero-Calvo, {\'A}lvaro and Strafford, David and Kamire, Rebecca and Childers, Amanda and Yates, Stephen F. and Rahislic, Emir and Smoke, Jason and Zheng, Sheng-Hai and Cameron, Peter and Thompson, Paul and et al.},
	year         = 2024,
	booktitle    = {Advances in Optical and Mechanical Technologies for Telescopes and Instrumentation VI},
	address      = {Yokohama, Japan},
	pages        = {131007H},
    publisher = {SPIE},
	doi          = {10.1117/12.3020718}
}

@book{Wu2013,
   author = {Zhizheng Wu and Azhar Iqbal and Foued Ben Amara},
   city = {Berlin, Heidelberg},
   doi = {10.1007/978-3-642-32229-7},
   edition = {1},
   isbn = {978-3-642-32228-0},
   publisher = {Springer},
   title = {Modeling and Control of Magnetic Fluid Deformable Mirrors for Adaptive Optics Systems},
   year = {2013}
}

@article{karbalaei_thermocapillarity_2016,
  author    = {Karbalaei, Alireza and Kumar, Ranganathan and Cho, Hyoung Jin},
  title     = {Thermocapillarity in Microfluidics-A Review},
  journal   = {Micromachines},
  volume    = {7},
  number    = {1},
  pages     = {13},
  year      = {2016},
  doi       = {10.3390/mi7010013},
}

@inproceedings{Gregg2024_FluidicMirrorSupportStructure,
  title        = {Structural Requirements and Scaling Analysis of a Fluidic Mirror Space Telescope Support Structure},
  author       = {Gregg, Christine and Balaban, Edward},
  booktitle    = {2024 IEEE Aerospace Conference (AERO)},
  year         = {2024},
  address      = {Big Sky, Montana, USA},
  pages        = {1--10},
  publisher    = {IEEE},
  doi          = {10.1109/AERO58975.2024.10520999},
}

@article{Gabay2025_FluidDynamicsLiquidMirror,
	title   = {Fluid dynamics of a liquid mirror space telescope},
	author  = {Gabay, Israel and Luria, Omer and Balaban, Edward and Gat, Amir D. and Bercovici, Moran},
	journal = {arXiv},
	year    = {2025},
	doi     = {10.48550/arXiv.2507.02812}
}

@article{Comstock2025,
  title        = {On the feasibility of spherical magnetic liquid mirror telescopes},
  author       = {Comstock, Eric A. and Chen, Hugh and Hu, Tianyang and Romero-Calvo, {\'A}lvaro},
  journal      = {Acta Astronautica},
  volume       = {230},
  pages        = {30--38},
  year         = {2025},
  doi          = {10.1016/j.actaastro.2025.01.066}
}

@misc{Balaban2025FLUTE_Poster,
  author       = {Balaban, Edward},
  title        = {Fluidic Telescope (FLUTE): Enabling the Next Generation of Large Space Observatories},
  howpublished = {NIAC Phase II Symposium, NASA Ames Research Center, USA},
  year         = {2025},
  note         = {Poster presentation},
  url          = {https://www.nasa.gov/wp-content/uploads/2025/09/niac-poster-balaban-flute-tagged.pdf}
}

@article{dimarco2009,
    author  = {Di Marco, P. and Grassi, W.},
    title   = {Effect of force fields on pool boiling flow patterns in normal and reduced gravity},
    journal = {Heat and Mass Transfer},
    volume  = {45},
    year    = {2009},
    pages   = {959--966},
    doi     = {10.1007/s00231-007-0328-6}
}

@article{Oron1997ThinFilms,
  author  = {Oron, Alexander and Davis, Stephen H. and Bankoff, S. George},
  title   = {Long-scale evolution of thin liquid films},
  journal = {Reviews of Modern Physics},
  volume  = {69},
  number  = {3},
  pages   = {931--980},
  year    = {1997},
  doi     = {10.1103/RevModPhys.69.931}
}

@book{Leal2007AdvancedTransport,
  author    = {Leal, L. Gary},
  title     = {Advanced Transport Phenomena},
  subtitle  = {Fluid Mechanics and Convective Transport Processes},
  publisher = {Cambridge University Press},
  address   = {Cambridge},
  year      = {2007},
  isbn      = {9780521839653}
}

@book{BornWolf1999,
  author    = {Born, Max and Wolf, Emil},
  title     = {Principles of Optics},
  publisher = {Cambridge University Press},
  address   = {Cambridge, United Kingdom},
  year      = {1999},
  chapter   = {1},
  pages     = {1--53},
  isbn      = {9780521642224},
  doi       = {10.1017/CBO9781139644181}
}

@article{Mahajan1983,
  author = {Mahajan, V. N.},
  title = {Strehl ratio for primary aberrations in terms of their aberration variance},
  journal = {Journal of the Optical Society of America},
  year = {1983},
  volume = {73},
  number = {6},
  pages = {860--861},
  doi = {10.1364/JOSA.73.000860}
}

@article{Carlson1995,
  author  = {Carlson, B. C.},
  title   = {Numerical Computation of Real or Complex Elliptic Integrals},
  journal = {Numerical Algorithms},
  volume  = {10},
  year    = {1995},
  pages   = {13--26},
  doi     = {10.1007/BF02198293}
}

@article{Plevachuk2014,
  author    = {Plevachuk, Yuriy and Sklyarchuk, Vasyl and Eckert, Sven and Gerbeth, Gunter and Novakovic, Rada},
  title     = {Thermophysical Properties of the Liquid {Ga}--{In}--{Sn} Eutectic Alloy},
  journal   = {Journal of Chemical \& Engineering Data},
  year      = {2014},
  volume    = {59},
  number    = {3},
  pages     = {757--763},
  doi       = {10.1021/je400882q},
  publisher = {American Chemical Society}
}

@article{Haskett2005,
  author    = {Haskett, Ryan P. and Witelski, Thomas P. and Sur, Jeanman},
  title     = {Localized Marangoni forcing in driven thin films},
  journal   = {Physica D: Nonlinear Phenomena},
  volume    = {209},
  number    = {1--4},
  pages     = {117--134},
  year      = {2005},
  issn      = {0167-2789},
  doi       = {10.1016/j.physd.2005.06.019},
  publisher = {Elsevier}
}

@article{Craster2009,
  author  = {Craster, R. V. and Matar, O. K.},
  title   = {Dynamics and stability of thin liquid films},
  journal = {Reviews of Modern Physics},
  volume  = {81},
  number  = {3},
  pages   = {1131--1198},
  year    = {2009},
  doi     = {10.1103/RevModPhys.81.1131}
}

@book {Rosensweig,
title = {Ferrohydrodynamics},
editor = {Rosensweig, Ronald E. [Verfasser/in]},
series = {Cambridge monographs on mechanics and applied mechanics},
address = {Cambridge},
publisher = {Cambridge Univ. Pr.},
year = {1985},
edition = {1. publ.},
isbn = {0521256240},
}

@inproceedings{Dietrich_RomeroCalvo_2026,
  title        = {Equilibrium and Stability of Orbiting Liquid Mirror Telescopes},
  author       = {Janoah Dietrich and {\'A}lvaro Romero-Calvo},
  booktitle    = {AIAA SciTech 2026 Forum},
  year         = {2026},
  address      = {Orlando, FL},
  publisher    = {American Institute of Aeronautics and Astronautics},
  doi          = {10.2514/6.2026-0312},
  note         = {Paper No. AIAA 2026-0312},
  pages        = {1--18}
}

@article{Kang2016,
  author    = {Kang, D. and Nadim, A. and Chugunova, M.},
  title     = {Dynamics and equilibria of thin viscous coating films 
               on a rotating sphere},
  journal   = {Journal of Fluid Mechanics},
  volume    = {791},
  pages     = {495--518},
  year      = {2016},
  publisher = {Cambridge University Press},
  doi       = {10.1017/jfm.2016.67}
}

@article{Gardner2023_JWST,
  author  = {Gardner, Jonathan P. and Mather, John C. and Abbott, Randy
             and Abell, James S. and Abernathy, Mark and others},
  title   = {The {James Webb Space Telescope} Mission},
  journal = {Publications of the Astronomical Society of the Pacific},
  year    = {2023},
  volume  = {135},
  number  = {1048},
  pages   = {068001},
  doi     = {10.1088/1538-3873/acd1b5}
}

@article{Krejci2018_FEEP,
  author  = {Krejci, David and Lozano, Paulo},
  title   = {Space Propulsion Technology for Small Spacecraft},
  journal = {Proceedings of the {IEEE}},
  year    = {2018},
  volume  = {106},
  number  = {3},
  pages   = {362--378},
  doi     = {10.1109/JPROC.2017.2778747}
}

@article{Wilson2023,
  author    = {Wilson, Lynn B. {III} and Salem, Chadi S. and Bonnell, John W.},
  title     = {Spacecraft Floating Potential Measurements for the {Wind} Spacecraft},
  journal   = {The Astrophysical Journal Supplement Series},
  year      = {2023},
  volume    = {269},
  number    = {2},
  pages     = {52},
  doi       = {10.3847/1538-4365/ad0633}
}

@article{Smirnov2024,
  author    = {Smirnov, S. and Podivilov, E. and Sturman, B.},
  title     = {Electrostatic conductive disc singularity resolved},
  journal   = {Journal of Applied Physics},
  year      = {2024},
  volume    = {135},
  pages     = {124301},
  doi       = {10.1063/5.0190593}
}

@book{Landau1984,
  author    = {Landau, L. D. and Lifshitz, E. M.},
  title     = {Electrodynamics of Continuous Media},
  edition   = {2nd},
  publisher = {Pergamon Press},
  address   = {Oxford},
  year      = {1984},
  series    = {Course of Theoretical Physics},
  volume    = {8}
}

@book{Jackson1999,
  author    = {Jackson, John David},
  title     = {Classical Electrodynamics},
  edition   = {3rd},
  publisher = {John Wiley \& Sons},
  address   = {New York},
  year      = {1999}
}

@article{Troy2003,
  author    = {Troy, Mitchell and Chanan, Gary},
  title     = {Diffraction effects from giant segmented-mirror telescopes},
  journal   = {Applied Optics},
  volume    = {42},
  number    = {19},
  pages     = {3745--3753},
  year      = {2003},
  doi       = {10.1364/AO.42.003745},
}

@article{Gardner2006,
  author    = {Gardner, Jonathan P. and Mather, John C. and Clampin, Mark
               and Doyon, Rene and Greenhouse, Matthew A. and Hammel,
               Heidi B. and Hutchings, John B. and Jakobsen, Peter and
               Lilly, Simon J. and Long, Knox S. and Lunine, Jonathan I.
               and McCaughrean, Mark J. and Mountain, Matt and Nella, John
               and Rieke, George H. and Rieke, Marcia J. and Rix, Hans-Walter
               and Smith, Eric P. and Sonneborn, George and Stiavelli, Massimo
               and Stockman, H. S. and Windhorst, Rogier A. and Wright,
               Gillian S.},
  title     = {The {James Webb Space Telescope}},
  journal   = {Space Science Reviews},
  volume    = {123},
  number    = {4},
  pages     = {485--606},
  year      = {2006},
  doi       = {10.1007/s11214-006-8315-7},
}

@article{Chanan2000,
  author    = {Chanan, Gary and Ohara, Catherine and Troy, Mitchell},
  title     = {Phasing the mirror segments of the {Keck} telescopes~{II}:
               the narrow-band phasing algorithm},
  journal   = {Applied Optics},
  volume    = {39},
  number    = {25},
  pages     = {4706--4714},
  year      = {2000},
  doi       = {10.1364/AO.39.004706},
}

@article{Menzel2023,
  author    = {Menzel, M. and Davis, M. and Parrish, K. and Lawrence, J. 
               and Stewart, A. and Cooper, J. and Irish, S. and Mosier, G. 
               and Levine, M. and Pitman, J. and Walsh, G. and Maghami, P. 
               and Thomson, S. and Wooldridge, E. and Boucarut, R. and 
               Feinberg, L. and Turner, G. and Kalia, P. and Bowers, C.},
  title     = {The Design, Verification, and Performance of the {James Webb Space Telescope}},
  journal   = {Publications of the Astronomical Society of the Pacific},
  volume    = {135},
  pages     = {058002},
  year      = {2023},
  doi       = {10.1088/1538-3873/acbb9f}
}

@inproceedings{Dichmann2014,
  author    = {Dichmann, Donald J. and Alberding, Cassandra M. and Yu, Wayne H.},
  title     = {Stationkeeping {Monte Carlo} Simulation for the {James Webb Space Telescope}},
  booktitle = {Proceedings of the International Symposium on Space Flight Dynamics 2014},
  address   = {Laurel, MD, USA},
  month     = {may},
  year      = {2014},
  note      = {NASA Technical Report {GSFC-E-DAA-TN14095}},
  url       = {https://ntrs.nasa.gov/citations/20140007519},
  pages     = {1--21}
}

@inproceedings{Hill1995HypervelocityAngle,
  author    = {Hill, David C. and Rose, M. Frank and Best, Steve R. and
               Crumpler, Michael S. and Crawford, Gary D. and
               Zee, Ralph H.-C. and Bozack, Michael J.},
  title     = {The Effect of Impact Angle on Craters Formed by
               Hypervelocity Particles},
  booktitle = {{LDEF}: 69 Months in Space. Third Post-Retrieval
               Symposium, Part 1},
  publisher = {NASA Langley Research Center},
  year      = {1995},
  month     = feb,
  url       = {https://ntrs.nasa.gov/citations/19950017419},
  pages     = {483--497}
}

@techreport{crawford1993hypervelocity,
  author      = {Crawford, Gary and Hill, David and Rose, Frank E. and
                 Zee, Ralph and Best, Steve and Crumpler, Mike},
  title       = {Hypervelocity Impact Study: {The} Effect of Impact
                 Angle on Crater Morphology},
  institution = {Space Power Institute, Auburn University},
  year        = {1993},
  type        = {Contractor Report},
  number      = {NASA-CR-192711},
  month       = mar
}

@book{marsh1980lasl,
  author    = {Marsh, Stanley P.},
  title     = {{LASL} Shock Hugoniot Data},
  publisher = {University of California Press},
  address   = {Berkeley},
  year      = {1980}
}

@article{birkhoff1948explosives,
  author  = {Birkhoff, Garrett and MacDougall, D. P. and Pugh, E. M. and Taylor, Geoffrey},
  title   = {Explosives with lined cavities},
  journal = {Journal of Applied Physics},
  year    = {1948},
  volume  = {19},
  number  = {6},
  pages   = {563--582},
  doi     = {10.1063/1.1698173}
}

@article{taylor1995meteoroid,
  author  = {Taylor, A. D.},
  title   = {The {Harvard} Radio Meteor Project velocity distribution reappraised},
  journal = {Icarus},
  year    = {1995},
  volume  = {116},
  number  = {1},
  pages   = {154--158},
  doi     = {10.1006/icar.1995.1117}
}

@book{melosh1989impact,
  author    = {Melosh, H. Jay},
  title     = {Impact Cratering: A Geologic Process},
  publisher = {Oxford University Press},
  address   = {New York},
  year      = {1989},
  series    = {Oxford Monographs on Geology and Geophysics},
  volume    = {11}
}

@article{grun1985collisional,
  author  = {Gr{\"u}n, E. and Zook, H. A. and Fechtig, H. and Giese, R. H.},
  title   = {Collisional balance of the meteoritic complex},
  journal = {Icarus},
  year    = {1985},
  volume  = {62},
  number  = {2},
  pages   = {244--272},
  doi     = {10.1016/0019-1035(85)90121-6}
}

@article{Fletcher2015,
  author  = {Fletcher, Alex and Close, Sigrid and Mathias, Donovan},
  title   = {Simulating plasma production from hypervelocity impacts},
  journal = {Physics of Plasmas},
  volume  = {22},
  number  = {9},
  pages   = {093504},
  year    = {2015},
  doi     = {10.1063/1.4930281}
}

@article{leinert1998,
  author  = {Leinert, Ch. and Bowyer, S. and Haikala, L. K. and Hanner, M. S.
             and Hauser, M. G. and Levasseur-Regourd, A.-C. and Mann, I.
             and Mattila, K. and Reach, W. T. and Schlosser, W. and others},
  title   = {The 1997 reference of diffuse night sky brightness},
  journal = {Astronomy and Astrophysics Supplement Series},
  year    = {1998},
  volume  = {127},
  pages   = {1--99},
  doi     = {10.1051/aas:1998105}
}

@article{rogalski2005,
  author  = {Rogalski, A.},
  title   = {{HgCdTe} infrared detector material: history, status and outlook},
  journal = {Reports on Progress in Physics},
  year    = {2005},
  volume  = {68},
  number  = {10},
  pages   = {2267--2336},
  doi     = {10.1088/0034-4885/68/10/R01}
}

@article{wright2015,
  author  = {Wright, G. S. and Wright, D. and Goodson, G. B. and Rieke, G. H.
             and Aitink-Kroes, G. and Alexov, A. and Libralato, M. and others},
  title   = {The {Mid-Infrared Instrument} for the {James Webb Space Telescope},
             {II}: Design and Build},
  journal = {Publications of the Astronomical Society of the Pacific},
  year    = {2015},
  volume  = {127},
  number  = {953},
  pages   = {595--611},
  doi     = {10.1086/682253}
}

@article{werner2004,
  author  = {Werner, M. W. and Roellig, T. L. and Low, F. J. and Rieke, G. H.
             and Rieke, M. and Hoffmann, W. F. and Young, E. and Houck, J. R.
             and Brandl, B. and Fazio, G. G. and others},
  title   = {The {Spitzer Space Telescope} Mission},
  journal = {The Astrophysical Journal Supplement Series},
  year    = {2004},
  volume  = {154},
  number  = {1},
  pages   = {1--9},
  doi     = {10.1086/422992}
}

@article{pilbratt2010,
  author  = {Pilbratt, G. L. and Riedinger, J. R. and Passvogel, T. and Crone, G.
             and Doyle, D. and Gageur, U. and Heras, A. M. and Jewell, C.
             and Metcalfe, L. and Ott, S. and Schmidt, M.},
  title   = {{Herschel Space Observatory}: An {ESA} facility for far-infrared
             and submillimetre astronomy},
  journal = {Astronomy and Astrophysics},
  year    = {2010},
  volume  = {518},
  pages   = {L1},
  doi     = {10.1051/0004-6361/201014759}
}

@article{rigby2023,
  author  = {Rigby, Jane R. and Lightsey, Paul A. and
             {Garc\'ia Mar\'in}, Macarena and others},
  title   = {How Dark the Sky: {The} {JWST} Backgrounds},
  journal = {Publications of the Astronomical Society of the Pacific},
  volume  = {135},
  number  = {1046},
  pages   = {048002},
  year    = {2023},
  doi     = {10.1088/1538-3873/acbcf4}
}

@article{ressler2015,
  author  = {Ressler, M. E. and Sukhatme, K. G. and Franklin, B. R. and others},
  title   = {The Mid-Infrared Instrument for the {James Webb Space Telescope}, {VIII}: The {MIRI} Focal Plane System},
  journal = {Publications of the Astronomical Society of the Pacific},
  volume  = {127},
  pages   = {675--692},
  year    = {2015},
  doi     = {10.1086/682258}
}

@article{poglitsch2010,
  author  = {Poglitsch, A. and Waelkens, C. and Geis, N. and others},
  title   = {The {Photodetector Array Camera and Spectrometer} ({PACS}) on the {Herschel Space Observatory}},
  journal = {Astronomy \& Astrophysics},
  volume  = {518},
  pages   = {L2},
  year    = {2010},
  doi     = {10.1051/0004-6361/201014535}
}

@book{rieke2003,
  author    = {Rieke, G. H.},
  title     = {Detection of Light: From the Ultraviolet to the Submillimeter},
  edition   = {2nd},
  publisher = {Cambridge University Press},
  year      = {2003},
  doi       = {10.1017/CBO9780511606496}
}

@inproceedings{Brooks2021,
  author    = {Brooks, Thomas E. and Stahl, H. Philip},
  title     = {Extreme dimensional stability thermal control test},
  booktitle = {Proc.\ SPIE Optical Engineering $+$ Applications},
  volume    = {11819},
  pages     = {118190Q},
  year      = {2021},
  doi       = {10.1117/12.2593969},
}

@article{Brooks2022,
  author  = {Brooks, Thomas E. and Stahl, H. Philip and Arnold, William R.
             and Perrygo, Charles and Greenhouse, Matthew A.},
  title   = {Precision thermal control technology to enable thermally stable telescopes},
  journal = {Journal of Astronomical Telescopes, Instruments, and Systems},
  volume  = {8},
  number  = {2},
  pages   = {024001},
  year    = {2022},
  doi     = {10.1117/1.JATIS.8.2.024001},
}

@inproceedings{Havey2019,
  author    = {Havey, Keith},
  title     = {Challenges and benefits to achieving sub-millikelvin thermal control
               stability on large space observatories},
  booktitle = {Proc.\ SPIE UV/Optical/IR Space Telescopes and Instruments},
  volume    = {11116},
  pages     = {111162G},
  year      = {2019},
  doi       = {10.1117/12.2534564},
}

@article{Rayleigh1879,
  author  = {Rayleigh, Lord},
  title   = {On the capillary phenomena of jets},
  journal = {Proceedings of the Royal Society of London},
  volume  = {29},
  pages   = {71--97},
  year    = {1879},
  doi     = {10.1098/rspl.1879.0015}
}

@article{Eggers_2008,
doi = {10.1088/0034-4885/71/3/036601},
year = {2008},
month = {feb},
volume = {71},
number = {3},
pages = {036601},
author = {Eggers, Jens and Villermaux, Emmanuel},
title = {Physics of liquid jets},
journal = {Reports on Progress in Physics},
}

@book{Lamb1932,
  author    = {Lamb, Horace},
  title     = {Hydrodynamics},
  edition   = {6},
  publisher = {Cambridge University Press},
  year      = {1932},
  isbn      = {0521458684}
}

@article{Ohnesorge1936,
  author  = {von Ohnesorge, Wolfgang},
  title   = {Die {Bildung} von {Tropfen} an {D{\"u}sen} und die {Aufl{\"o}sung} fl{\"u}ssiger {Strahlen}},
  journal = {Zeitschrift f{\"u}r Angewandte Mathematik und Mechanik},
  volume  = {16},
  number  = {6},
  pages   = {355--358},
  year    = {1936},
  doi     = {10.1002/zamm.19360160611}
}

@misc{LUVOIR_2019,
  author       = {{The LUVOIR Team}},
  title        = {The {LUVOIR} Mission Concept Study Final Report},
  year         = {2019},
  eprint       = {1912.06219},
  archivePrefix = {arXiv},
  primaryClass = {astro-ph.IM},
  doi          = {10.48550/arXiv.1912.06219}
}

@book{Astro2020,
  author       = {{National Academies of Sciences, Engineering, and Medicine}},
  title        = {Pathways to Discovery in Astronomy and Astrophysics for the 2020s},
  year         = {2023},
  publisher    = {The National Academies Press},
  address      = {Washington, {DC}},
  doi          = {10.17226/26141}
}

@article{MacFarlane2002,
  author    = {MacFarlane, D. R. and Forsyth, S. A. and Golding, J. and Deacon, G. B.},
  title     = {Ionic liquids based on imidazolium, ammonium and pyrrolidinium salts of the dicyanamide anion},
  journal   = {Green Chemistry},
  year      = {2002},
  volume    = {4},
  number    = {5},
  pages     = {444--448},
  doi       = {10.1039/B205641K}
}

@article{HandschuhWang2021,
author = {Handschuh-Wang, Stephan and Stadler, Florian J. and Zhou, Xuechang},
title = {Critical Review on the Physical Properties of Gallium-Based Liquid Metals and Selected Pathways for Their Alteration},
journal = {The Journal of Physical Chemistry C},
volume = {125},
number = {37},
pages = {20113-20142},
year = {2021},
doi = {10.1021/acs.jpcc.1c05859},
}

@article{Kramer2014,
  author    = {Kramer, Rebecca K. and Boley, J. William and Stone, Howard A.
               and Weaver, James C. and Wood, Robert J.},
  title     = {Effect of Microtextured Surface Topography on the Wetting
               Behavior of Eutectic Gallium--Indium Alloys},
  journal   = {Langmuir},
  year      = {2014},
  volume    = {30},
  number    = {2},
  pages     = {533--539},
  doi       = {10.1021/la404356r},
}

@manual{SpaceX2020Starship,
  author       = {{Space Exploration Technologies Corp.}},
  title        = {Starship Users Guide},
  edition      = {Revision 1.0},
  month        = mar,
  year         = {2020},
  url          = {https://www.spacex.com/media/starship_users_guide_v1.pdf},
}

@manual{MATLAB2023,
  author       = {{The MathWorks Inc.}},
  title        = {{MATLAB version: 9.15.0 (R2023b)}},
  year         = {2023},
  organization = {The MathWorks, Inc.},
  address      = {Natick, Massachusetts, United States},
  url          = {https://www.mathworks.com}
}

@article{MelcherTaylor1969,
  author  = {Melcher, J. R. and Taylor, G. I.},
  title   = {Electrohydrodynamics: A review of the role of interfacial shear stresses},
  journal = {Annual Review of Fluid Mechanics},
  volume  = {1}, 
  pages = {111--146}, 
  year = {1969},
  doi     = {10.1146/annurev.fl.01.010169.000551}
}

@article{Borgnis1953,
  author  = {Borgnis, F. E.},
  title   = {Acoustic radiation pressure of plane compressional waves},
  journal = {Reviews of Modern Physics},
  volume  = {25}, 
  number = {3}, 
  pages = {653--664}, 
  year = {1953},
  doi     = {10.1103/RevModPhys.25.653}
}

@article{Beyer1978,
  author  = {Beyer, Robert T.},
  title   = {Radiation pressure---the history of a mislabeled tensor},
  journal = {The Journal of the Acoustical Society of America},
  volume  = {63}, 
  number = {4}, 
  pages = {1025--1030}, 
  year = {1978},
  doi     = {10.1121/1.381833}
}

@article{Nagakura2025,
  author  = {Nagakura, Koki and Fushimi, Tatsuki and Tsutsui, Ayaka and Ochiai, Yoichi},
  title   = {Dynamic caustics by ultrasonically modulated liquid surface},
  journal = {Scientific Reports},
  volume  = {15}, 
  number = {31928}, 
  year = {2025},
  doi     = {10.1038/s41598-025-16190-3}
}

@article{Wang2017,
  author  = {Wang, Z. H. and Wang, S. D. and Meng, X. and Ni, M. J.},
  title   = {{UDV} measurements of single bubble rising in a liquid metal {Galinstan}
             with a transverse magnetic field},
  journal = {International Journal of Multiphase Flow},
  volume  = {94}, 
  pages = {201--208}, 
  year = {2017},
  doi     = {10.1016/j.ijmultiphaseflow.2017.05.001}
}

@article{Li2021,
  author  = {Li, Zhengwei and Xu, Zhiwu and Zhao, Degang and Chen, Shu and Yan, Jiuchun},
  title   = {Ultrasonic cavitation at liquid/solid interface in a thin {Ga--In}
             liquid layer with free surface},
  journal = {Ultrasonics Sonochemistry},
  volume  = {71}, 
  pages = {105356}, 
  year = {2021},
  doi     = {10.1016/j.ultsonch.2020.105356}
}

\end{document}